\documentclass[aip,jcp,reprint]{revtex4-1}

\usepackage{amsmath}
\usepackage{amsfonts}
\usepackage{amssymb}
\usepackage{graphicx}
\usepackage{subscript}

\usepackage[utf8x]{inputenc}
\usepackage{color}
\usepackage{bbm}
\usepackage{bm}

\usepackage{comment}

\begin{document}

\title{Equilibrium structures of anisometric, quadrupolar particles 
confined to a monolayer}

\author{Thomas Heinemann$^\star$}
\email{thomasheinemann@mailbox.tu-berlin.de} 
\affiliation{Institut f\"ur Theoretische Physik, Technische
  Universit\"at Berlin, Hardenbergstra{\ss}e 36, D-10623 Berlin,
  Germany}

\author{Moritz Antlanger$^\star$} 
\email{moritz.antlanger@tuwien.ac.at}
\affiliation{\mbox{Institut f\"ur Theoretische Physik, TU Wien, Wiedner Hauptstra{\ss}e 8-10, A-1040 Wien, Austria}}
  
\author{Martial Mazars}
\email{martial.mazars@th.u-psud.fr}
\affiliation{Laboratoire de Physique Th\'eorique (UMR 8627), CNRS, Univ. Paris-Sud, Universit\'e Paris-Saclay, 91405 Orsay, France}

\author{Sabine H.L. Klapp}
\email{klapp@physik.tu-berlin.de}
\affiliation{Institut f\"ur Theoretische Physik, Technische
  Universit\"at Berlin, Hardenbergstra{\ss}e 36, D-10623 Berlin,
  Germany}
  
\author{Gerhard Kahl}
\email{gerhard.kahl@tuwien.ac.at}
\affiliation{Institut f\"ur Theoretische Physik and Center for
  Computational Materials Science (CMS), TU Wien, Wiedner
  Hauptstra{\ss}e 8-10, A-1040 Wien, Austria}

$^\star$authors who have contributed equally to the research
\begin{abstract} 
We investigate the structural properties of a two-dimensional system
of ellipsoidal particles carrying a linear quadrupole moment in their
center.  These particles represent a simple model for a variety of
uncharged, non-polar conjugated organic molecules.  Using optimization
tools based on ideas of evolutionary algorithms, we first examine the
ground state structures as we vary the aspect ratio of the particles
and the pressure.  Interestingly, we find, besides the intuitively
expected T-like configurations, a variety of complex structures,
characterized with up to three different particle orientations. In an
effort to explore the impact of thermal fluctuations, we perform
constant-pressure molecular dynamics simulations within a range of
rather low temperatures. We observe that ground state structures
formed by particles with a large aspect ratio are in particular suited
to withstand fluctuations up to rather high temperatures. Our
comprehensive investigations allow for a deeper understanding of
molecular or colloidal monolayer arrangements under the influence of a
typical electrostatic interaction on a coarse-grained level.

\end{abstract}

\maketitle





\section{Introduction}
\label{sec:introduction}

In recent years there is increasing interest in understanding the film
morphologies of anisotropic, conjugated organic molecules at inorganic
surfaces.  A prime example are systems of optically active molecules
at inorganic surfaces; these so-called hybrid inorganic/organic
systems (HIOS) represent a very promising material class in
optoelectronics.\cite{blumstengel_2008,blumstengel_2010,draxl_2014,schlesinger_2015}
Typically, the corresponding organic molecules are strongly
anisotropic in shape and are characterized by complex charge
distributions, often dominated by a quadrupole
moment.~\cite{kleppmann_2015} By manipulating the orientational
structure of such organic layers it is possible to tune the efficiency
of the charge carrier transport~\cite{yanagi_1997} and thus to
optimize the efficiency of the hybrid system.  Somewhat earlier,
organic molecules at interfaces have attracted attention in the
context of so-called Langmuir monolayers, that is, two-dimensional
(2D) films of typically amphiphilic molecules constrained to a
liquid-gas interface.~\cite{kaganer_1993,kaganer_1999} Again, the
orientational and translational structure in such systems can be quite
complex.

From the theoretical side, a full microscopic treatment of an organic
molecular layer is still challenging, and this holds particularly for
HIOS (where the substrate is typically patterned).  On the level of
atomistic molecular dynamics (MD) simulations, one has to consider a
large number of interactions such as atomic bonding, van der Waals
(Lennard-Jones) potentials, contributions due to bending or torsion of
a molecule (see e.g. Ref.~\onlinecite{wang_2004}), and electrostatic
interactions. Even then, it often turns out that atomistic MD
calculations do not correctly predict the arrangement of several
particles, one main reason being the insufficient treatment of
polarizability effects.~\cite{sinnokrot_2002} An example
that demonstrates this issue is the quadrupolar molecule
benzene, where a representation via atomic point
charges~\cite{pettersson_1987} or multipoles~\cite{price_1987} fails
to reproduce the corresponding dimer configuration. This suggests to
employ quantum chemical approaches to evaluate the electronic
structure of the entire system, instead. However, at the
moment, this level of complexity is computationally far
too demanding, especially at finite temperatures.

Given these difficulties, the goal of the present paper is to
understand {\em generic} aspects of the structural behavior of
quadrupolar, anisotropic molecules at interfaces based on a {\em
  coarse-grained} model.  Specifically, we investigate a 2D
many-particle system composed of ellipse-shaped particles with an
embedded, axially symmetric quadrupole tensor (higher-order multipoles are
neglected).  The quadrupole is oriented along one of the ellipsoidal
axes as it is the case, e.g., in benzene~\cite{golubkov_2006} or
naphtalene, whose quadrupole tensor nearly has axial symmetry.~\cite{chablo_1981} We further assume that the center of
the particles is confined to a 2D plane, and that the molecules are
allowed to rotate only within this plane.

Interestingly, 2D quadrupolar systems are, so far, rather
unexplored. This contrasts the situation in the 3D case, where a
considerable amount of pioneering work involving quadrupolar particles
at finite
temperatures~\cite{bates_1998,boublik_1990,vega_1992,neal_1999,neal_2000}
and in the ground state \cite{miller_2008} is available.  On the other
hand, there are a number of simulation studies dealing with
non-quadrupolar, shape-anisotropic particles in 2D, examples being
hard
ellipsoids,~\cite{bautistacarbajal_2014,vieillardbaron_1972,cuesta_1989,cuesta_1990}
rectangles~\cite{martinezraton_2009} or hard spherocylinders.\cite{bates_2000}
In particular, the density-driven
nematic-isotropic transition of hard ellipsoids was already
investigated in Ref.~\onlinecite{cuesta_1990} and turned out being
continuous or of first order depending on the aspect ratio. We also
mention a density functional study~\cite{kaganer_1998} of 2D systems
of anisotropic particles exhibiting isotropic phases and phases with
simultaneous orientational and translational order. Thus, the
important open question to be explored in the present article concerns
the interplay of shape anisotropy and quadrupolar intermolecular
interactions.

To this end we employ two types of numerical calculations, that is,
optimization techniques based on ideas of evolutionary algorithms (EA)
and constant-pressure MD simulations, revealing the system's behavior
in the ground state and at finite temperatures.  To facilitate the
calculations, the non-electrostatic part of the interaction between
two molecules is modeled via a purely repulsive, relatively stiff (but
not infinitely hard) pair potential that foots on the ellipse-like
contact distance suggested by Berne and Pechukas.~\cite{berne_1972}
The analytical form of this pair potential resembles an anisotropic
soft-sphere potential, but has a higher exponent.\cite{sarman_2000}
This potential turned out to be suitable for both type of numerical
calculations. Van der Waals interactions between the molecules are
entirely neglected. Although this may seem somewhat unrealistic for
real molecules, there are several advantages of the resulting
(repulsive) potential: it reduces the number of parameters, it allows
(as we will demonstrate) for a transferability between results at
different system parameters, and it makes it possible to establish a
connection of the present study to investigations of anisotropic
colloidal particles in confined geometry.\cite{musevic_2006,skarabot_2008,ognysta_2009}

The remainder of this article is organized as follows.  In
Sec.~\ref{sec:model} we introduce the model and the pair interactions
in detail. Section~\ref{sec:methods_data_analysis} is devoted to the
methods used to calculate equilibrium structures (at $T=0$ and $T>0$)
and corresponding data analysis methods.  The resulting ground state
structures at different pressures and aspect ratios are presented in
Sec.~\ref{subsec:ground_state}, and the corresponding
finite-temperature results are discussed in
Sec.~\ref{subsec:finite_temperatures}.  Finally, we summarize our
findings in Sec.~\ref{sec:conclusion}. Appendices A to C provide
additional, complementary information.

\section{Model}
\label{sec:model}

In our investigations we consider a two-dimensional system, embedded
in the $(x, y)$-plane, where the anisotropic particles are only
allowed to rotate around the $z$-axis. We use reduced units
throughout, introducing parameters $\sigma_0$ for the length scale, $\epsilon_0$ for the energy scale, and $m_0$ for the unit of mass.

\begin{figure*}[htbp]
\begin{center}
\includegraphics[width=15cm]{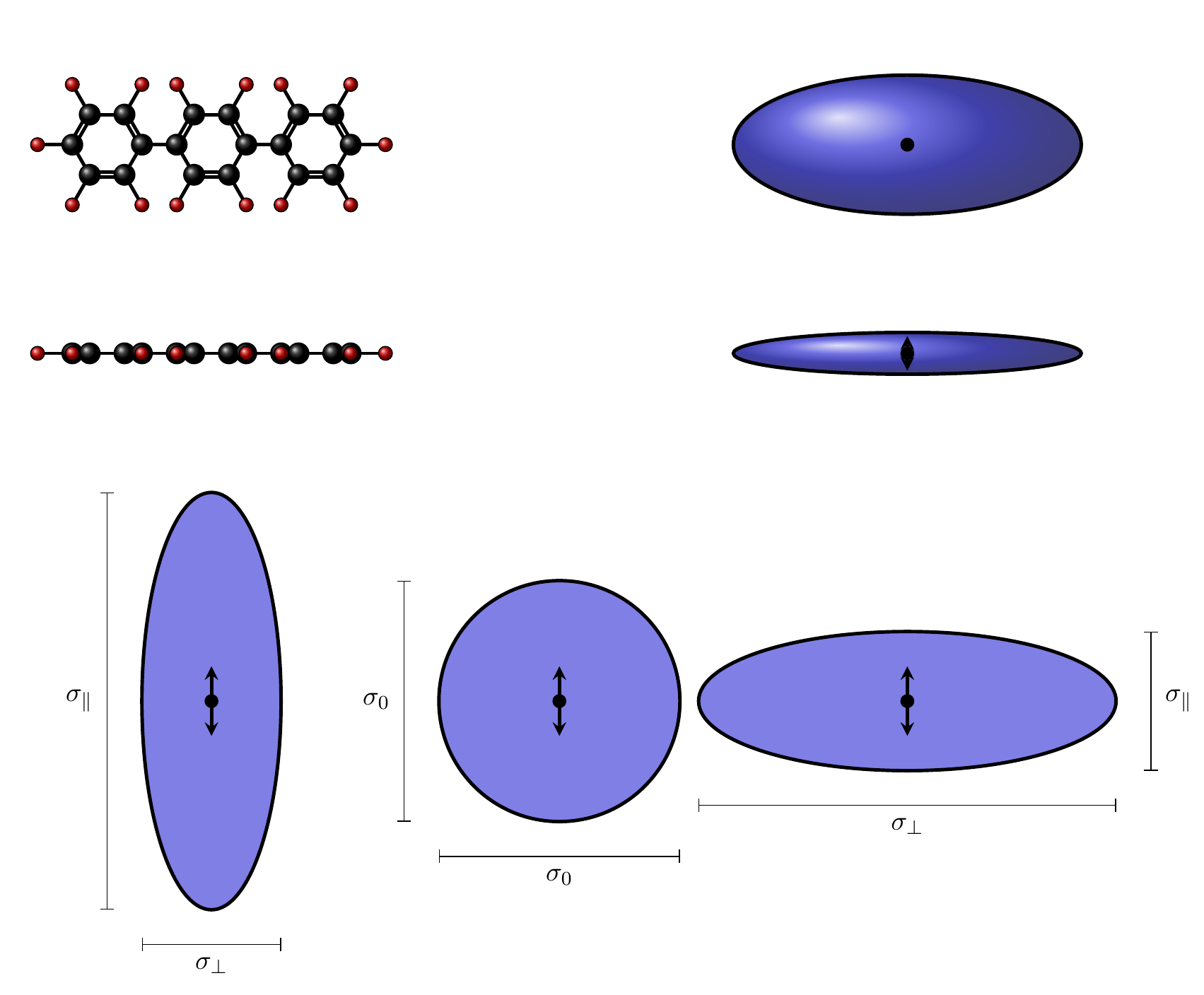}
\caption{Taken from Ref.~\onlinecite{antlanger_phd_2015}. Top-left panel: top and side views of a simple organic
  molecule (shown here is para-terphenyl (C$_{18}$H$_{14}$), which is composed of three benzene rings.  Molecules of this type
  show a strong anisometry and feature a complicated charge
  distribution. Top-right panel: simple organic molecules (as shown in
  the top-left panel) are modeled in this contribution as soft
  ellipsoids with an embedded quadrupole moment. Due to the high
  degree of symmetry of the underlying model, the quadrupole moment is
  assumed to be linear. Bottom panels: schematic view of our soft,
  anisometric particles carrying a linear quadrupole moment
  (visualized by the double-arrow) with $\kappa>1$ (left panel),
  $\kappa=1$ (center panel), and $\kappa<1$ (right panel). $\sigma_0$,
  $\sigma_{\parallel}$, and $\sigma_{\perp}$ are defined in the text.}
\label{fig:model}
\end{center}
\end{figure*}
Our particles are assumed to have an elliptic shape, characterized by
the lengths of the two main axes, $\sigma_{\parallel}$ and
$\sigma_{\perp}$; the indices refer to the orientation of the
corresponding axes relative to the linear quadrupolar moment to be
introduced below (see Figure \ref{fig:model}). Defining a
dimensionless shape anisotropy parameter, $\kappa$, we impose via
$$ 
\sigma_{\parallel} = \sigma_0 \sqrt{\kappa} ~~~~ {\rm and} ~~~~
\sigma_{\perp} = \frac{\sigma_0}{\sqrt{\kappa}} 
$$
that the surface area of our particles is independent of the actual
value of $\kappa$. The isometric case of a disc is recovered for
$\kappa = 1$.

The interaction between two-particles (with indices $i$ and $j$) can
be split into a short-ranged, anisotropic repulsion, $V_{\rm sr}({\bf
  r}_{ij}, \hat {\bf u}_i, \hat {\bf u}_j)$, and a long-ranged
potential, $V_{\rm lr}({\bf r}_{ij}, \hat {\bf u}_i, \hat {\bf u}_j)$;
the latter one stems from the interaction of the linear quadrupoles
that the particles carry. We define the distance $r_{ij} = | {\bf r}_i
- {\bf r}_j|$ between the centers of the particles and further $\hat
{\bf r}_{ij} = {\bf r}_{ij}/r_{ij}$. The $\hat {\bf u}_i$ and $\hat
{\bf u}_j$ are the normalized orientation vectors of the respective
linear quadrupolar moments. Within our model the shape of the particle
is the same for $\kappa$ and $1/\kappa$, while the respective
orientations of the embedded quadrupole moment are perpendicular for
the cases $\kappa$ and $1/\kappa$ (see bottom panels of Figure
\ref{fig:model}). Consequently, by choosing values of $\kappa$ both
smaller and larger than unity we are able to consider these two
orientations of the quadrupolar moment within the same model.

With the above vectors at hand we introduce
\begin{equation}
a = \hat{{\bf u}}_i \cdot \hat{{\bf r}}_{ij} ~~~~
b = \hat{{\bf u}}_j \cdot \hat{{\bf r}}_{ij} ~~~~
c = \hat{{\bf u}}_i \cdot \hat{{\bf u}}_j .
\end{equation}

For the short-range contribution of the inter-particle interaction
(subscript ``sr'') we use an anisotropic, repulsive potential: it has
the simple functional form of an inverse power law (IPL) interaction,
while its dependence on the connecting vector, ${\bf r}_{ij}$, and on
the orientations of the quadrupolar moments, ${\bf \hat{u}}_i$ and
${\bf \hat{u}}_j$, is inspired by a Gay-Berne potential:\cite{gay_1981}
\begin{eqnarray}
V_{\rm sr}({\bf r}_{ij}, \hat{\bf u}_i, \hat{\bf u}_j) & = 
& 4 \epsilon({\bf r}_{ij}, \hat{\bf u}_i, \hat{\bf u}_j) 
\left[ \frac{\sigma({\bf r}_{ij}, \hat{\bf u}_i, \hat{\bf u}_j)}{r_{ij}} 
\right]^{18} ;
\label{eqn:Vsr}
\end{eqnarray}
here
\begin{eqnarray}
\epsilon({\bf r}_{ij}, \hat{\bf u}_i, \hat{\bf u}_j) & = & \epsilon_0 = 
{\rm const.,} \\
\sigma({\bf r}_{ij}, \hat{\bf u}_i, \hat{\bf u}_j) & = & 
\sigma_{\perp} \left\{ 1-\frac{\chi}{2} 
\left[ \frac{\left( a+b \right)^2}{1+\chi c} + 
\frac{\left( a-b \right)^2}{1-\chi c} \right] \right\}^{-1/2} \label{eqn:sigma}\\
\chi & = & \frac{\kappa^2-1}{\kappa^2+1} .
\end{eqnarray}
The electrostatic part of the inter-particle interaction is based on
{\it linear} quadrupole moments, for which the quadrupole
tensor,~\cite{gray_1984} $\underline{\hat{\bf Q}}$, becomes in its
eigenbasis
\begin{align}
\underline{\hat{\bf Q}} &= 
\left( \begin{array}{c c c} Q_{xx} & 0 & 0 \\ 
                            0 & Q_{yy} & 0 \\ 
                            0 & 0 & -Q_{xx}-Q_{yy} 
\end{array} \right) \nonumber
\\
&= Q \left( \begin{array}{c c c} -1/2 & 0 & 0 \\ 
                                    0 & -1/2 & 0 \\ 
                                    0 & 0 & 1 \end{array} \right) ,
\end{align}
with $Q$ being the strength of the moment.

The interaction between two linear quadrupole moments can be written
as \cite{gray_1984,boublik_1990}
\begin{multline}
V_{\rm lr}(Q^2, {\bf r}_{ij}, \hat{{\bf u}}_i, \hat{{\bf u}}_j) = 
\frac{1}{4\pi\varepsilon_{\rm pm}} \frac{3}{4} \frac{Q^2}{r_{ij}^5} \times
\\
\times\left[ 1 - 5 a^2 - 5 b^2 - 15 a^2 b^2 +2 \left( c - 5 a b \right)^2 \right] ,
\label{eqn:Vlr}
\end{multline}
with $a$, $b$, and $c$ defined above and $\varepsilon_{\rm pm}$ being
the vacuum permittivity. Even though the interaction decays slower
with the distance (i.e., $\sim 1/r^5$) than, e.g., van der Waals
interactions ($\sim 1/r^6$) we can still calculate inter-particle
energies with sufficient accuracy via a real-space lattice sum. Here
we have used a cutoff radius $R_{\rm cut}/\sigma_0=30$ for both,
electrostatic and repulsive interactions.
\begin{figure}[htbp]
\begin{center}
\includegraphics[width=5cm]{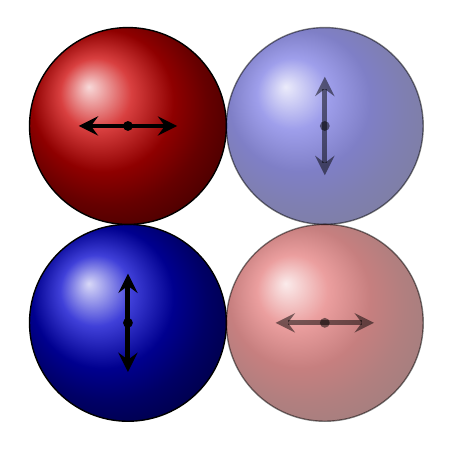}
\caption{Typical arrangement of {\it isolated} quadrupolar particles
  with weakly anisometric shape (i.e., $\kappa$ close to unity),
  induced by the fact that neighboring particles tend to arrange in a
  mutually orthogonal orientation.}
\label{fig:t_configuration}
\end{center}
\end{figure}
The interaction of two linear quadrupoles depends on their respective
orientations. In contrast to dipoles, which tend to line up
head-to-tail, isolated linear quadrupoles prefer the so-called
T-configuration for $\kappa$ close to unity (see Figure
\ref{fig:t_configuration}). A more detailed discussion of the impact
of shape anisotropy on the two-particle arrangement with a minimal
electrostatic energy is given in Appendix~\ref{sec:Two-particle
  arrangement with minimal electrostatic energy}.
    
Throughout, our numerical calculations were performed in the
$NPT$-ensemble, where the relevant thermodynamic potential for vanishing temperature is the
enthalpy, $H$,
\begin{eqnarray}
H & = & E + P S_0 ;
\end{eqnarray}
here $E$ is the internal energy, $P$ is the pressure, and $S_0$ is the
area occupied by the system. Further, we introduce the temperature of
the system, $T$.

Based on the scales of length ($\sigma_0$), mass ($m_0$), and energy
($\epsilon_0$), the quadrupole strength, $Q^2$, the pressure $P$, the temperature $T$ and the time $t$
are expressed via their respective reduced units, $(Q^*)^2 = Q^2/(4
\pi \varepsilon_{\rm pm} \sigma_0^5 \epsilon_0)$, $P^* = P
\sigma_0^2/\epsilon_0$, $T^*= k_{\rm B} T/\epsilon_0$ (with $k_{\rm B}$ being the Boltzmann constant) and $t^*=t/\sqrt{\sigma_0^2 m_0/\epsilon_0}$.
		
\section{Methods and data analysis}
\label{sec:methods_data_analysis}

In our approach we investigate in a {\it first} step the self-assembly
scenarios of our particles at {\it zero} temperature (i.e., the ground
state configurations). This information helps us to classify
archetypical particle arrangements (each characterized by a particular
spatial and orientational order), which, in turn, help us to
understand the strategy of the particles of how to assemble in the
energetically most favorable manner. For these investigations we
employ an optimization tool based on evolutionary algorithms, detailed
below.

In a {\it second} step we take advantage of these ground state
configurations and use them as starting configurations in molecular
dynamics (MD) simulations, performed at small, but {\it finite}
temperatures. These investigations provide information on the
thermodynamic stability of the ordered phases.

\subsection{Evolutionary algorithms}
\label{subsec:EA}

In our effort to identify ground state configurations of our systems,
we use an optimization tool based on ideas of evolutionary algorithms
(EAs).~\cite{gottwald_2005} EAs are heuristic approaches designed to
search for global minima in high-dimensional search spaces and for
problems that are characterized by rugged energy landscapes.

In an effort to be compatible with the requirements of an
$NPT$-ensemble, we introduce in our approach a unit cell of variable
area and shape which creates (together with its periodic images) a
system of infinite extent. In the desired configuration (the so-called
ground state configuration) particles are located and oriented in this
cell in such a way as to minimize the internal energy of the system,
which at vanishing temperature is equivalent to the enthalpy.

We initialize the algorithm by creating a set of configurations where
particles are located in the cell at random positions and have random
orientations. These arrangements are graded by their respective fitness
value, a quantity that provides evidence on how suitable this
configuration is to solve the optimization problem. Since we are
interested in finding ground state structures, a high fitness value of
a particular configuration corresponds to a low value of the enthalpy
per particle.

Then we iteratively use existing particle arrangements to create new
ones by applying one of the two following operations: crossover and
mutation. In the former one we first select two configurations where
the choice is biased by high fitness values of the two
configurations. Characteristic features of both particle arrangements
(such as lattice vectors, particle positions, and particle
orientations) are then combined to form a new configuration. The
mutation operation, on the other hand, introduces random changes to a
randomly chosen configuration, such as moving or rotating an
arbitrarily chosen particle or changing the lattice vectors. Typically
2000 iterations are required for a particular state point until 
convergence towards the global minimum has been achieved.

Our implementation of EAs is memetic, i.e., we combine global and
local search techniques: each time a new configuration has been
created with one of the two above mentioned EA operations, we apply
the L-BFGS-B \cite{byrd_1995} algorithm which guides us to the nearest
local minimum.

This algorithm has been applied successfully for a broad spectrum of
systems, both in two and three dimensions (see, for instance, Refs.~\onlinecite{fornleitner_2008,fornleitner_2009,doppelbauer_2010,doppelbauer_2012,bianchi_2012,noya_2014}). Within the context of
the present contribution, it should also be mentioned, that a
suitable extension of the formalism is also able to cope with long-ranged interactions, as they are encountered in charged systems (see Refs.~\onlinecite{antlanger_2014,antlanger_2015}).

For the present contribution we have performed with our EA approach
computations for 211 evenly-spaced values of $\kappa \in [0.4, 2.5]$
and for several combinations of $(Q^*)^2 \in \{ 0.2, 2, 20\}$ and $P^*
\in \{ 0.1, 1, 10\}$. Due to computational limitations we have
considered unit cells that contain up to twelve particles.

\subsection{Molecular dynamics simulations}
\label{subsec:MD}

In order to investigate structural changes of our system at finite
temperature, we perform molecular dynamics (MD) simulations at
constant pressure and temperature for an ensemble of $N=840$
particles.  To this end we use the Berendsen weak-coupling
scheme,~\cite{berendsen_1984} where the classical Newtonian equations
of motion are supplemented by terms representing the coupling to a
heat- and a pressure-bath.  

In the following, we first introduce the translational equations of
motion proposed by Berendsen {\it et al.},~\cite{berendsen_1984}
specializing to the case of a two-dimensional system.  Considering a
particle with index $i$ ($i=1,\ldots,N)$ one has
\begin{align}
\dot{\mathbf{r}}_i&=
\mathbf{v}_i+\frac{K }{ \tau_{\rm P}} 
(\underline{\mathcal{P}}-\underline{\mathbf{P}}) \mathbf{r}_i\text{,}
\label{eqn:translational eqm}
\\
m_0 \dot{\mathbf{v}}_i&=
\mathbf{F}_i + m_0 \frac{1}{2\tau_{\rm trans}} 
\left( \frac{T}{\mathcal{T}_{\rm trans}}-1\right)\mathbf{v}_i\text{,}
\\
\dot{\mathbf{b}}_\alpha&=
\left[\frac{K }{ \tau_{\rm P}} 
(\underline{\mathcal{P}}-\underline{\mathbf{P}} ) \right]
\mathbf{b}_\alpha .
\label{eqn:eqm_box_vectors}
\end{align}
where the vectors $\mathbf{b}_\alpha$ ($\alpha = 1, 2$) define the
simulation box and its shape.

In the above equations, the dot represents a time-derivative. Further,
${\bf r}_i$ and ${\bf v}_i$ denote the position and the velocity of
particle with index $i$, respectively; ${\mathbf{F}}_i$ is the force
on particle $i$ with mass $m_0$, and $K$ is the compressibility. The
pressure tensors, $\underline {\mathcal P}$ (actual pressure) and
$\underline {\bf P}$ (target pressure), are defined below. Further,
$\mathcal{T}_{\rm trans}$ is the actual kinetic temperature, $k_{\rm
  B} \mathcal{T}_{\rm trans}=\sum_i m_i v_i^2/[2(N-1)]$ where the
denominator $2(N-1)$ represents the $2N$ translational degrees of freedom minus two constraints due to momentum conservation in each spatial dimension.
To ensure this conservation during the numerical
integration, we set the total momentum to zero every 100 time steps.

The (time) constants $\tau_{\rm trans}$ and $\tau_{\rm P}$ determine
the speed of relaxation of $\mathcal{T}_{\rm trans}$ and $\underline
{\mathcal P}$ towards their respective target values defined by the
bath. Specifically, the corresponding coupling equations read
\begin{align}
\dot{\mathcal{T}}_{\rm trans}=
\frac{T-\mathcal{T}_{\rm trans}}{\tau_{\rm trans}} \qquad 
\dot{\underline{\mathcal{P}}}=
\frac{\underline{\mathbf{P}}-{\underline {\mathcal P}}}{\tau_{\rm P}} ,
\label{eqn:T and P control}
\end{align}
where the pressure tensors are defined as
\begin{align}
&\underline{\mathbf{\mathcal P}} = \frac{1}{2 S_0} \left[\sum_i m_0
  \mathbf{v}_i \otimes \mathbf{v}_i+ \sum_{i, j; i<j} \mathbf{r}_{ij}
  \otimes \mathbf{F}_{ij} \right]
  \\
  &\text{and}\nonumber
  \\
  &\underline{\mathbf{P}}=
\frac{1}{2} P\,\,\underline{\mathbf{1}} .
\label{eqn:pressure tensors}
\end{align}
Here, $\otimes$ denotes the dyadic product of two vectors, $S_0$
stands for the area of the simulation cell, $\underline {\bf 1}$ is
the unit-tensor, and ${\bf F}_{ij}$ is the force on particle $i$ exerted
by particle $j$. 

To describe the rotational motion of the particles,
Eqs.~\eqref{eqn:translational eqm}--\eqref{eqn:pressure tensors} are
supplemented with the following differential equations for the angular
velocity, $\boldsymbol{\omega}_i$, of particle $i$:
\begin{align}
\ddot{\hat{\mathbf{u}}}_i & =
\dot{\boldsymbol{\omega}}_i\times \hat{\mathbf{u}}_i+
\boldsymbol{\omega}_i \times \dot{\hat{\mathbf{u}}}_i\text{,}
\label{eqn:eqm_omega1}
\\
I \dot{\boldsymbol{\omega}}_i & = 
\boldsymbol{M}_i + \frac{I}{2\tau_{\rm rot}} 
\left( \frac{T}{\mathcal{T}_{\rm rot}}-1\right)\,\boldsymbol{\omega}_i
\label{eqn:eqm_omega2} .
\end{align}            
Here, $\boldsymbol{M}_i$ and $I$ are the torque on particle $i$ and
the corresponding moment of inertia of particle $i$, respectively;
$\times$ denotes the vector product. The kinetic temperature for the
rotation is defined as $k_{\rm B} \mathcal{T}_{\rm rot}=\sum_i I
\omega_i^2/N$.  This temperature is controlled in analogy to
Eqs.~\eqref{eqn:T and P control}.

To solve the translational equations of motion, i.e. Eqs.~\eqref{eqn:translational eqm}--\eqref{eqn:pressure tensors} we
use a modified leap-frog integrator as proposed in
Ref.~\onlinecite{berendsen_1984}. For the solution of the rotational equations of motion,
Eqs.~\eqref{eqn:eqm_omega1}--\eqref{eqn:eqm_omega2}, an analogous
procedure is performed (see Ref.~\onlinecite{fincham_1993}).
             
To initialize the particle positions in our system, we use the unit
cells obtained from the ground state calculations in the preceding
investigations (see Subsec.~\ref{subsec:EA}). Specifically, we take
for each ($\kappa$-dependent) ground state the respective unit cell
and arrange copies of this cell such that the simulation cell has a
minimal circumference. For this simulation cell (which is now defined
by the simulation box vectors $\mathbf{b}_1$ and $\mathbf{b}_2$)
periodic boundary conditions are applied.
The pair forces $\mathbf{F}_{ij}$ and torques $\mathbf{M}_{ij}$ are
truncated at $R_{\rm cut}=6\sigma_0$ and are then shifted in order to
make them vanish smoothly.

The MD calculations have been performed at several values of the
reduced temperature $T^*$, that is,
$T^{*}=0.1, 0.2,\dots, 1.6$.
In our simulations we used a time increment $\Delta t^*=0.00136172$
(corresponding at a macroscopic level to $\Delta t\approx
2~\textrm{fs}$). An adequate choice of the time step is imposed by the
mass of the particles, for which we have assumed the mass of benzene,
$m_0=78 {\rm u}$.  Each simulation extends over 210~000 time
steps. During the first 20~000 MD steps, the target temperature is
gradually increased from $T^*=0$ towards the respective target
value. Structural quantities are then extracted only during the final
10~000 time steps of the simulation. The values for the temperature-
and pressure-coupling constants $\tau_{\rm trans}$, $\tau_{\rm rot}$
and $\tau_{\rm P}$ were chosen to be $80$ time steps. Of course, the
actual volume change of the box is also influenced by the
compressibility; its reduced, dimensionless counterpart, $K^*=K m_0^2
\sigma_0^2/\epsilon_0$, is set to 0.01. We assume the mass
distribution within the particles to be homogeneous and thus obtain
for the dimensionless moment of inertia $I^*=0.25 \left(
\kappa+\frac{1}{\kappa} \right)$.
       
\subsection{Structural analysis -- order parameters}
\label{subsec:order_parameters}

In order to quantify both the positional and the orientational order
of the particles we introduce three different sets of order
parameters:

\begin{itemize}
\item[(i)] The {\it positional} order can be quantified via
  two-dimensional bond-orientational order parameters (BOOPs),~\cite{strandburg_1988}
  introducing, in addition, weight factors~\cite{mickel_2013}
  that are related to the side lengths of the
  Voronoi polygon around each particle:
\begin{eqnarray}
\Psi_{n} & = & \left<\sum_{i=1}^N \frac{1}{\sum_{j \in \mathcal{N}_i} l_{ij}} 
\left| \sum_{j \in \mathcal{N}_i}   l_{ij} \exp(\imath n \phi_{ij}) \right|\right> ;
\label{eq:Psi}
\end{eqnarray}
here ($j \in \mathcal{N}_i$) indicates that particle $j$ is a nearest
neighbor of particle $i$ and $l_{ij}$ is the length of the side of the
Voronoi polygon that separates the two particles. $\phi_{ij}$ is the
angle enclosed by the bond between particles $i$ and $j$ and the
reference axis, which we choose as the $x$-axis. For the parameter $n$
we have chosen the values 4 and 6, highlighting thus via $\Psi_4$ or
$\Psi_6$ four- or six-fold symmetry, respectively. While BOOPs are
calculated for only the best configuration at vanishing temperature,
averages over several configurations are taken in MD simulations,
indicated by the brackets in Eq. \eqref{eq:Psi} and in the following.
	
\item[(ii)] An obvious candidate for quantifying the {\it
  orientational} order of the particles is the nematic order parameter
  $S$, which is introduced via the tensor order parameter
  $\underline{\bf T}$, defined as
\begin{eqnarray}
\underline{\bf T} & = & 2\left< \frac{1}{N} 
\sum_{i}\mathbf{\hat{u}}_{i} \otimes 
\mathbf{\hat{u}}_{i} \right> - \underline{\bf 1}
\nonumber .
\end{eqnarray}
$S$ is then simply the positive eigenvalue of the tracless tensor $\underline{\bf T}$,
which can easily be calculated as
\begin{eqnarray}
S & = & \sqrt{T_{xx}^2+T_{xy}^2}. \nonumber
\end{eqnarray}
The eigenvector associated with $S$ is called the director
$\hat{\bm{d}}$ and indicates a preferred orientation in the
configuration. Note that $S$ is always positive, except in the case
that all elements of $\underline{\bf T}$ vanish: then no preferred
direction is present in the system.
	
A different way of how to measure the orientational order is via the
parameter $\beta$, defined as 
\begin{eqnarray}
\beta &=& \left< \frac{1}{N}\sum_{i=1}^N \frac{1}{\sum_{j \in \mathcal{N}_i} l_{ij}} 
\sum_{j \in \mathcal{N}_i} l_{ij}\left| 
(\mathbf{\hat{u}}_{i}\cdot \mathbf{\hat{u}}_j)  \right| \right>  \nonumber .
\end{eqnarray}
\item[(iii)] In addition, a variety of order parameters combining {\it
  positional} and {\it orientational} order can be defined. An example
  of such an order parameter, which has been used in a previous
  contribution \cite{georgiou_2014} is
\begin{eqnarray}
\alpha\! &=&\!\left<\!\frac{1}{2N}\sum_{i=1}^N 
\frac{1}{\sum_{j \in \mathcal{N}_i} l_{ij}} \sum_{j \in \mathcal{N}_i} l_{ij}
\left| (\mathbf{\hat{u}}_{i}\cdot \mathbf{\hat{r}}_{ij})^2 \!+\! 
(\mathbf{\hat{u}}_{j}\cdot \mathbf{\hat{r}}_{ij})^2  \right|\! \right> \nonumber .
\end{eqnarray}
\end{itemize}
Concluding, we emphasize that orientational order refers to the
orientation of the {\it particles}; the orientation of the linear
quadrupolar moment is then imposed by the value of $\kappa$: if
$\kappa > 1$, then the orientation of the moment is parallel to the
main axis of the particles, while for $\kappa < 1$ these two
orientations are perpendicular to each other.

\section{Results} 
\label{sec:results}

\subsection{Ground state configurations}
\label{subsec:ground_state}

In our discussion of the ground state configurations we first present
the different structural archetypes that we could identify with our
EA-approach. We then present the diagram of states: it provides
information on the $\kappa$-ranges where the respective structures are
the energetically most stable ones; we further discuss how the
thermodynamic properties and the order parameters of our systems vary
over a representative range of $\kappa$.

\subsubsection{Structural archetypes}
\label{subsubsec:archetypes}

\begin{figure*}
\begin{center}
\includegraphics[width=15cm]{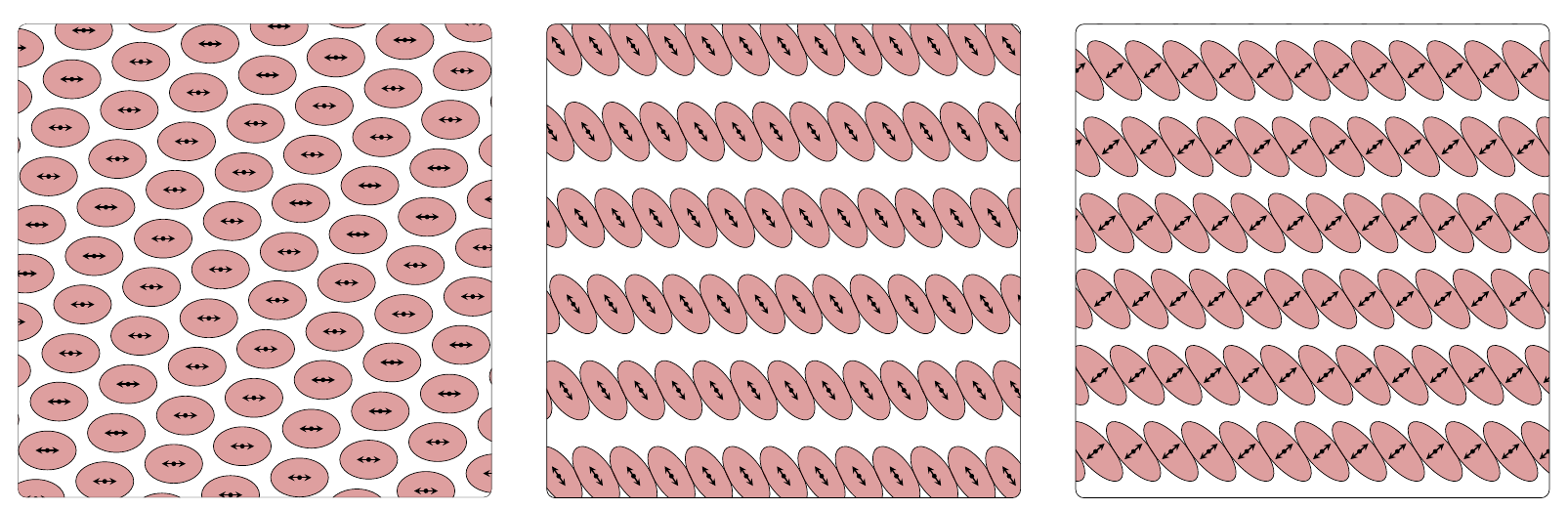}
\caption{Taken from Ref.~\onlinecite{antlanger_phd_2015}. Snapshots of ground state configurations with one preferred particle
  orientation. Left panel: P-configuration ($(Q^*)^2=0$, $P^*=1$,
  and $\kappa=1.5$). Center panel: PD-configuration ($(Q^*)^2=2$,
  $P^*=1$, and $\kappa=2.1$). Right panel: PD-configuration
  ($(Q^*)^2=2$, $P^*=1$, and $\kappa=0.4$).}
\label{fig:systems_quadrupolar_snap1}
\end{center}
\end{figure*}

\paragraph{Structures with one preferred orientation}
For $\kappa$-values both considerably larger or smaller than unity
(i.e., by a factor of $\approx 2$), the particles tend to align
parallel to each other in case of strong anisotropy: they accomplish
this by forming tilted rows, characterized by a high density of
particles along these lines; these arrangements are denoted as {\it
  parallel displaced}-configurations (``PD''; see Figure
\ref{fig:systems_quadrupolar_snap1}). Since quadrupolar particles
avoid a head-to-tail arrangement, neighboring rows repel each other,
inducing thereby large gaps between these lanes; the parallel offset
between neighboring rows can be very sensitive to small changes in
$\kappa$, leading to a slight modulation of the BOOPs as
  functions of $\kappa$ (to be discussed below). An interesting
special case of this structure is observed for a vanishing quadrupolar
moment (i.e., $(Q^*)^2=0$), where particles form a distorted hexagonal
lattice, denoted as the {\it parallel}-configuration (``P''; see top
left panel of Figure \ref{fig:systems_quadrupolar_snap1}).
\begin{figure*}
\begin{center}
\includegraphics[width=15cm]{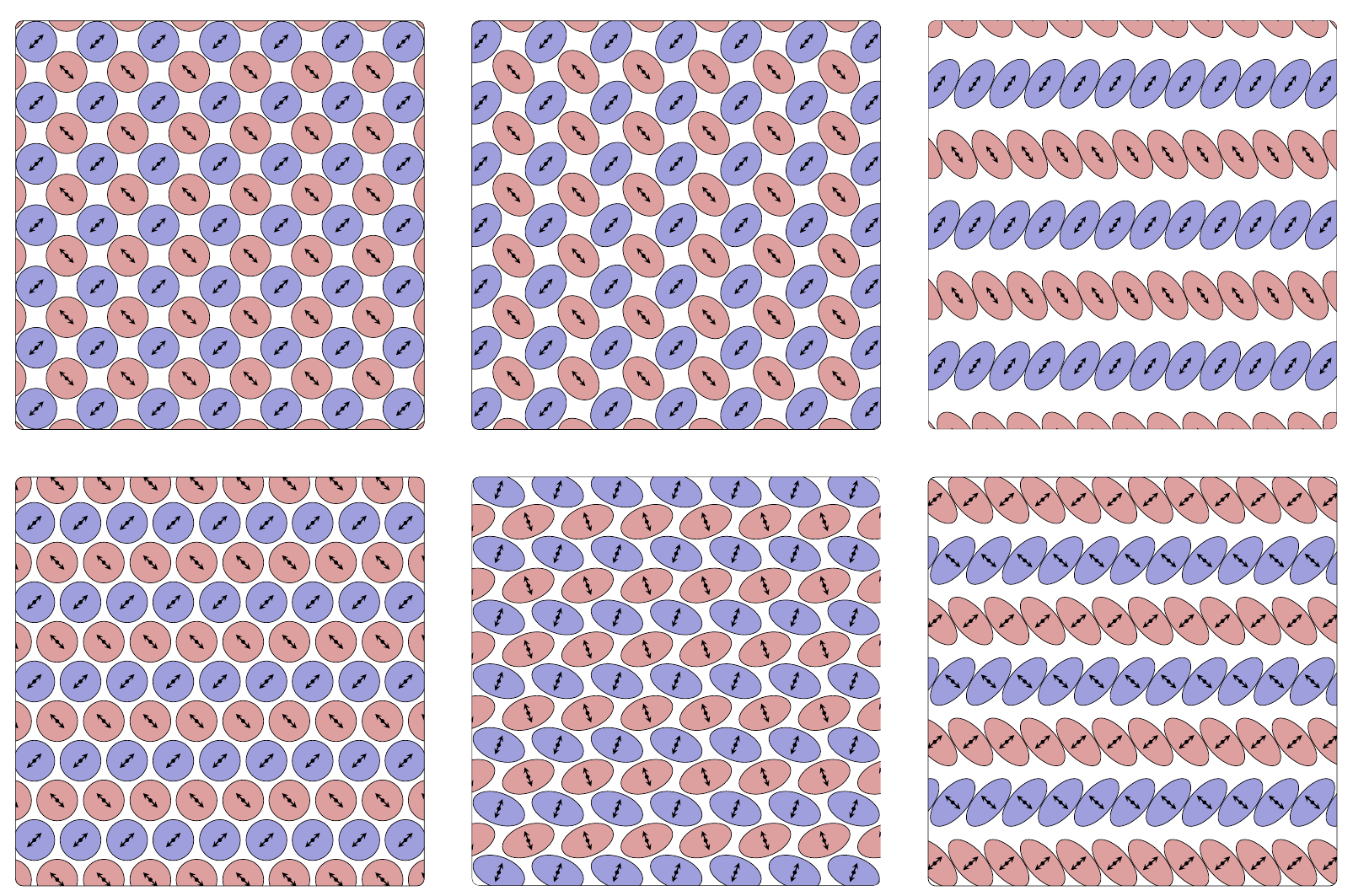}
\caption{Taken from Ref.~\onlinecite{antlanger_phd_2015}. Snapshots of ground state configurations with two preferred particle
  orientations. Different colors indicate different particle
  orientations. Left column: top panel -- T$_{\rm sq}$-configuration
  ($(Q^*)^2=2$, $P^*=1$, and $\kappa=1$), bottom panel -- T$_{\rm
    hex}$-configuration ($(Q^*)^2=0.2$, $P^*=10$, and
  $\kappa=1$). Center column: top panel -- HB$_{\rm
    dense}$-configuration ($(Q^*)^2=2$, $P^*=1$, and $\kappa=1.38$),
  bottom panel -- HB$_{\rm dense}$-configuration ($(Q^*)^2=2$,
  $P^*=1$, and $\kappa=0.58$). Right column: top panel -- HB$_{\rm
    loose}$-configuration ($(Q^*)^2=2$, $P^*=1$, and $\kappa=1.83$),
  bottom panel -- HB$_{\rm loose}$-configuration ($(Q^*)^2=2$,
  $P^*=1$, $\kappa=0.48$).}
\label{fig:systems_quadrupolar_snap2}
\end{center}
\end{figure*}
\paragraph{Structures with two preferred particle orientations}
Here the preferred structural arrangements are throughout {\it
  herringbone}-configurations (``HB'') where particles form
alternating, parallel lanes, each of them being characterized by a
specific particle orientation: in Figure
\ref{fig:systems_quadrupolar_snap2} particles pertaining to different
lanes are colored red and blue, respectively. The relative orientation
between neighboring particles depends in a highly sensitive manner on
$\kappa$.

Apart from two special cases (that are encountered for $\kappa$-values
very close to unity and which will be discussed below) two different
versions of the HB-configurations emerge from a closer analysis of the
obtained structures:
\begin{itemize}
\item[(i)] for $\kappa$-values somewhat closer to unity, we observe a
  rather dense structure, which we denote HB$_{\rm
    dense}$-configuration (see panels in the central column of Figure
  \ref{fig:systems_quadrupolar_snap2});
\item[(ii)] for $\kappa$-values that differ more strongly from
  unity, particles arrange in densely-populated rows of
  alternating orientation (see panels in the right column of Figure
  \ref{fig:systems_quadrupolar_snap2}). Since neighboring rows repel
  each other, this structure has a lower overall density which we
  therefore denote as HB$_{\rm loose}$.
\end{itemize}
The two special cases of the HB structure mentioned above are observed
for $\kappa$-values very close to unity where neighboring particles
prefer strict mutual orthogonal orientations with respect to their
nearest neighbors. (i) For large $(Q^*)^2$- and small $P^*$-values, a
perfect arrangement of mutually orthogonally oriented particles with
an underlying square pattern can be observed, denoted as the {\it
  square T}-configuration (``T$_{\rm sq}$''; see left panel of
Figure \ref{fig:systems_quadrupolar_snap2}). (ii) Further, a closely
related arrangement has been identified which is now based on an
underlying hexagonal pattern; it is denoted as the {\it hexagonal
  T}-configuration (``T$_{\rm hex}$''; see bottom left panel of Figure
\ref{fig:systems_quadrupolar_snap2}); for this particular
configuration a mutually perfect orthogonal orientation of nearest
neighbors can only be realized for $\kappa=1$.
\begin{figure*}
\begin{center}
\includegraphics[width=15cm]{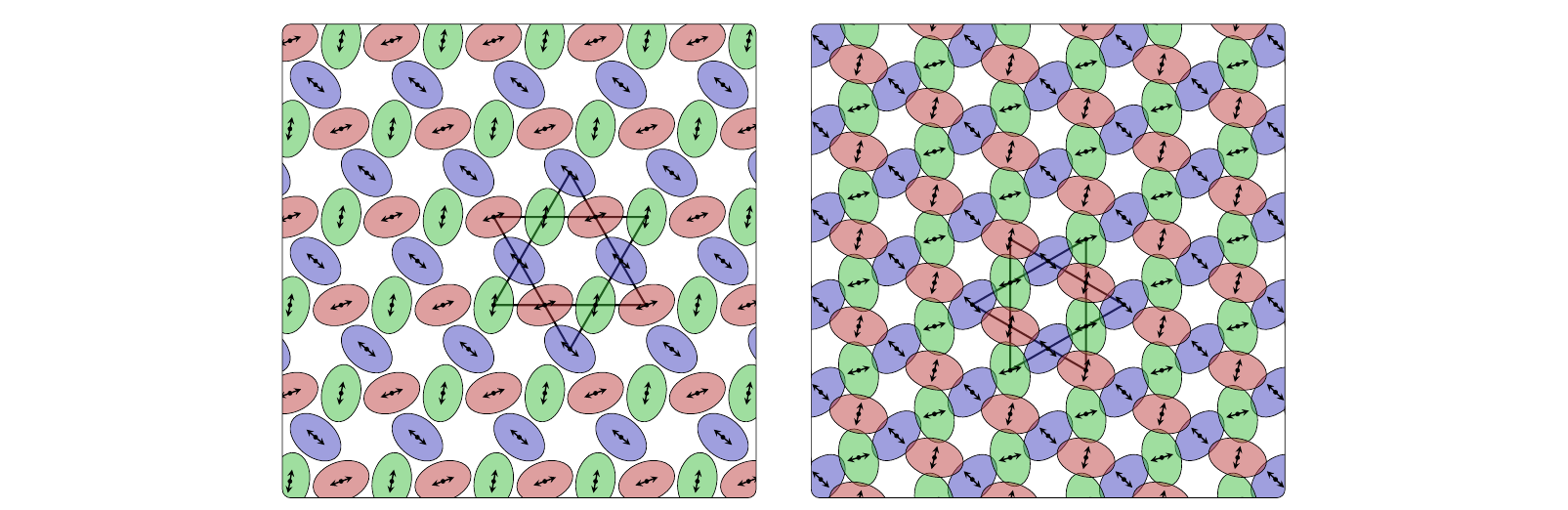}
\caption{Taken from Ref.~\onlinecite{antlanger_phd_2015}. Snapshots of ground state configurations with three preferred particle
  orientations. Different colors indicate different particle
  orientations. Left panel: TH-configuration ($(Q^*)^2=2$, $P^*=1$, and
  $\kappa=1.5$), right panel: TH-configuration ($(Q^*)^2=20$,
  $P^*=1$, and $\kappa=0.65$). Lines emphasize underlying triangles and hexagons.}
\label{fig:systems_quadrupolar_snap3}
\end{center}
\end{figure*}
\paragraph{Structures with three preferred particle orientations}
A very interesting phenomenon observed in our system is the occurrence
of structures where particles arrange in three preferred orientations,
characterized by a vanishing nematic order parameter $S$.

The {\it spatial} particle arrangement is here -- irrespective of the
values of $Q^*$ and $P^*$ -- reminiscent of a {\it trihexagonal}
tiling,~\cite{grunbaum_1987} consisting of regular hexagons that are
connected by triangles; we thus denote it as the ``TH''-configuration
(see Figure \ref{fig:systems_quadrupolar_snap3} where also the
hexagons and the triangles are highlighted).  Throughout, the relative
angle between the orientations of neighboring particles is $\pi/3$,
reflected in the order parameter which assumes a value of $\beta =
\cos^2(\pi/3) = 1/4$ (see Figure \ref{fig:ea_2_1}).

We note that the occurrence of the TH-configuration strongly depends
on the choice of $(Q^*)^2$ and $P^*$; in some cases, this particle
arrangement is not observed at all (see Figure \ref{fig:ea_02_01}). In
contrast, variants of the \textrm{HB}-configuration can almost always
be observed in some range of $\kappa$ (see discussion below).
\begin{figure*}
\begin{center}
\includegraphics[width=15cm]{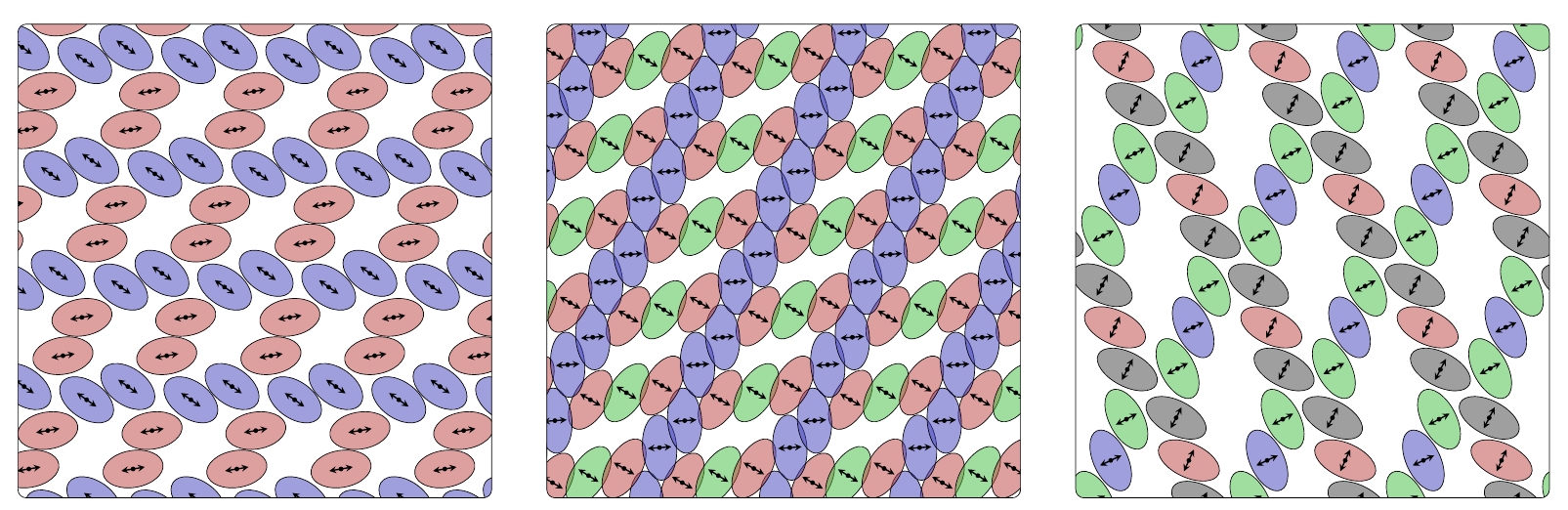}
\caption{Taken from Ref.~\onlinecite{antlanger_phd_2015}. Snapshots of more complicated ground state configurations, so-called
  B-configurations. Different colors indicate different particle
  orientations. Left panel: ($(Q^*)^2=2$, $P^*=1$, and $\kappa=1.66$),
  center panel: ($(Q^*)^2=20$, $P^*=10$, and $\kappa=0.52$), right
  panel: ($(Q^*)^2=2$, $P^*=0.1$, and $\kappa=0.53$).}
\label{fig:systems_quadrupolar_snap4}
\end{center}
\end{figure*}
\paragraph{More complicated structures}	
Using the EA-approach we also identify more complicated structures,
not conforming to the mechanisms described above. These structures
require a larger number of particles per cell and turn out to be
stable only within very small ranges of $\kappa$ (see
Subsec.\ \ref{subsubsec:diagram_states}). While we observe several
different variants of such particle configurations, they share a
common feature, namely a mesh-like lattice, where chains of particles
undulate back and forth (see Figure
\ref{fig:systems_quadrupolar_snap4}). We denote them as {\it branched}
structures (``B'').

\subsubsection{Diagram of states}
\label{subsubsec:diagram_states}

We now discuss the diagram of states which summarizes the occurrence
of the previously identified archetypes of ground state configurations
of our system for selected values of $(Q^*)^2$ and $P^*$ as we vary
$\kappa$. Since most of the interesting features of our systems can
already be captured in the diagram of states obtained for $(Q^*)^2=2$
and $P^*=1$, we focus from now onwards on this particular set of
parameters: we present data for the enthalpy and for a selection of
appropriate order parameters introduced in
Subsec.\ \ref{subsec:order_parameters}, that help to identify the
respective structures. The filling fraction $\eta$, which is also
displayed in the following figures, is defined as $\eta = N \pi
\sigma_0^2/4/S_0$.

At this point we remind the reader that the {\it shape} (and thus the
orientation of the particles) is invariant under the transformation
$\kappa \leftrightarrow 1/\kappa$ while the relative orientations of
the quadrupolar moments for the cases $\kappa$ and $1/\kappa$ are
mutually orthogonal: with this symmetry in mind, we can easily
disentangle the impact of $(Q^*)^2$ and $P^*$ on the structure
formation by comparing the relevant physical properties for the cases
$\kappa$ and $1/\kappa$.
\begin{figure*}
\includegraphics[width=0.8\textwidth]{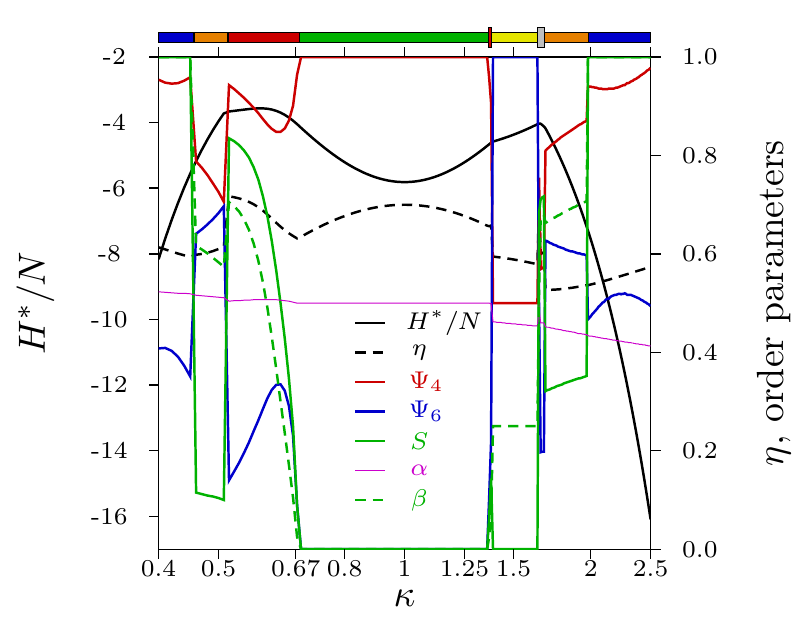}
\caption{Taken from Ref.~\onlinecite{antlanger_phd_2015}. Reduced enthalpy per particle ($H^*/N$; left vertical axis)
  as well as filling fraction $\eta$ (as defined in the text) and
  different order parameters (for their definitions see
  Subsec. \ref{subsec:order_parameters}) of the observed ground state
  configurations (right vertical axis) as functions of $\kappa$ (as
  labeled) for $(Q^*)^2=2$ and $P^*=1$. Note the logarithmic scale
  along the $\kappa$-axis. The horizontal bar above the panel
  specifies the $\kappa$-range where the respective ordered structural
  archetype is the energetically most favorable one via the following
  color code: T$_{\rm sq}$ (green), HB$_{\rm dense}$ (red), HB$_{\rm
    loose}$ (orange), PD (blue), TH (yellow), B
  (grey).}
\label{fig:ea_2_1}
\end{figure*}
These properties are displayed for $(Q^*)^2=2$ and $P^*=1$ in Figure
\ref{fig:ea_2_1} along with a color-coded, horizontal bar (above this
panel) which indicates these $\kappa$-ranges where the respective
particle configurations are the energetically most favorable ones.

On a qualitative level we observe that the enthalpy curve is
continuous over the entire investigated $\kappa$-range; however, it
shows kinks at particular values of the aspect ratio which provide a first
evidence for the occurrence of  discontinuous transitions between the ground state
configurations.  The specific enthalpy, $H^*/N$, shows a pronounced
local minimum at $\kappa = 1$ (i.e., for circular particle shapes):
here the T-configuration is dominant, verified by
the fact that in this $\kappa$-region $\Psi_4 = 1$ and $\Psi_6 = 0$. In
addition, we observe $S = 0$ and $\beta = 0=\cos^2 (\pi/2)$,
indicating thereby a relative orthogonal orientation of neighboring
particles.  In contrast, for more anisometric particle shapes (i.e.,
small and large $\kappa$-values), the enthalpy rapidly decays to
rather small values; actually $H$ tends to minus infinity for
$\kappa \to 0$ and $\kappa \to \infty$ as the quadrupoles start to
overlap and the repulsive soft core shrinks as $\sigma_{\perp} \to
0$. For small and large $\kappa$-values the formation of parallel rows
is energetically most favorable and PD-configurations are
observed. Here the nematic order parameter $S$ is the appropriate
quantity to characterize the emerging structure: indeed, $S$ assumes
in these $\kappa$-regions the value 1, indicating that a single
orientation prevails.

In contrast, for the two intermediate $\kappa$-ranges, where the
enthalpy assumes local maxima the ground state configurations for
$\kappa$ and $1/\kappa$ are distinctively different; for these
$\kappa$-values we observe the formation of more complex structures,
reflected by a rather intricate variation of the different order
parameters with $\kappa$: (i) {\it Increasing} first $\kappa$ beyond
the region where the T-configuration is stable, the system changes --
after a very narrow $\kappa$-region where the HB$_{\rm
  dense}$-structure occurs -- via a discontinuous {\it structural}
transition (identified by a jump in $\eta$) into the TH-configuration;
the latter one is characterized by $S = 0$, $\Psi_6 = 1$ and $\beta =
0.25 = \cos^2 (\pi/3)$ which provides evidence that the difference in
orientation angles between nearest neighbors is
$\pi/3$. Upon further increasing $\kappa$ we pass again a very
narrow interval where a branched structure is stable and we eventually
reach -- again via a discontinuous structural transition -- the
HB$_{\rm loose}$-structure; for this configuration none of the order
parameters assume any characteristic value. Eventually we identify for
even larger $\kappa$-values the aforementioned PD-structure. (ii) {\it
  Decreasing}, on the other hand, the value of $\kappa$ beyond the
range where the T-structure is stable, we observe HB-configurations:
first the one with the larger density (i.e., the HB$_{\rm
  dense}$-structure) and then the HB$_{\rm loose}$-structure. The
shape of $\beta$ (see Figure \ref{fig:ea_2_1}) indicates that the
transition between HB$_{\rm dense}$ and HB$_{\rm loose}$ (and also the
subsequent transition to the PD-structure) is of first order, i.e., at
some point the relative orientation of neighboring particles is
discontinuous.

At this point it should be mentioned that $\eta$ and the order parameters
do not necessarily behave in the same manner as the system changes from one
structure to the other: some order parameters or $\eta$ may change abruptly,
indicating a first order phase transition, while the other parameters change 
continuously. As an example we refer the reader to the transition HB$_{\rm loose}$ $\to$
PD for $\kappa \sim 0.45$. 

\begin{figure*}
\includegraphics[width=0.8\textwidth]{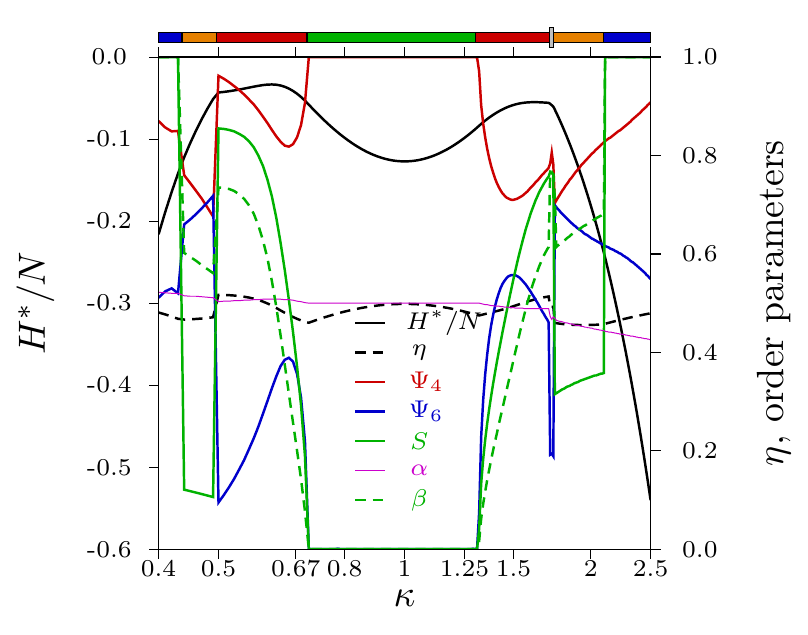}
\caption{
Taken from Ref.~\onlinecite{antlanger_phd_2015}. Reduced enthalpy per particle ($H^*/N$; left vertical axis)
  as well as filling fraction $\eta$ (as defined in the text) and
  different order parameters (for their definitions see
  Subsec. \ref{subsec:order_parameters}) of the observed ground state
  configurations (right vertical axis) as functions of $\kappa$ (as
  labeled) for $(Q^*)^2=0.2$ and $P^*=0.1$. Note the logarithmic scale
  along the $\kappa$-axis. The horizontal bar above the panel
  specifies the $\kappa$-range where the respective ordered structural
  archetype is the energetically most favorable one via the following
  color code: T$_{\rm sq}$ (green), HB$_{\rm dense}$ (red), HB$_{\rm
    loose}$ (orange), PD (blue), TH (yellow), B
  (grey).}
\label{fig:ea_02_01}
\end{figure*}

\begin{figure*}
\includegraphics[width=0.8\textwidth]{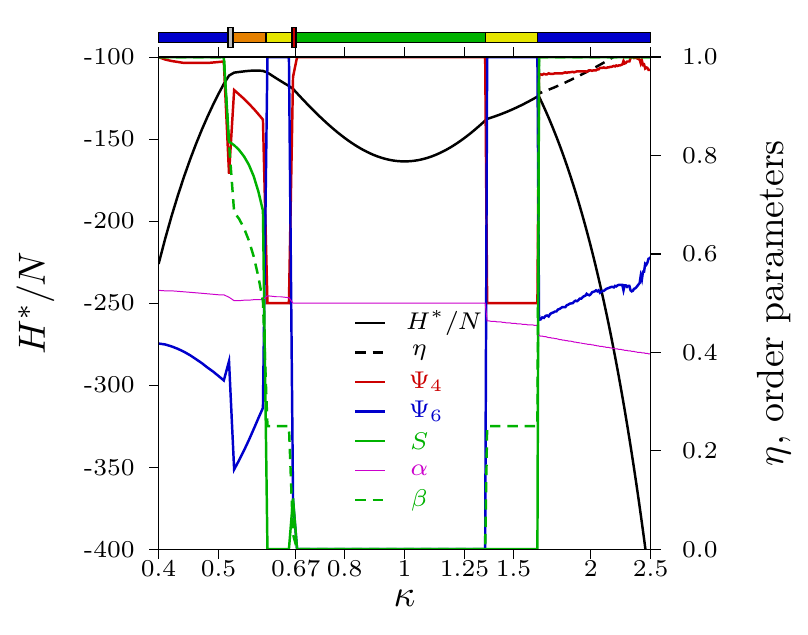}
\caption{(Color online)
Taken from Ref.~\onlinecite{antlanger_phd_2015}. Reduced enthalpy per particle ($H^*/N$; left vertical axis)
  as well as filling fraction $\eta$ (as defined in the text) and
  different order parameters (for their definitions see
  Subsec. \ref{subsec:order_parameters}) of the observed ground state
  configurations (right vertical axis) as functions of $\kappa$ (as
  labeled) for $(Q^*)^2=20$ and $P^*=10$. Note the logarithmic scale
  along the $\kappa$-axis. The horizontal bar above the panel
  specifies the $\kappa$-range where the respective ordered structural
  archetype is the energetically most favorable one via the following
  color code: T$_{\rm sq}$ (green), HB$_{\rm dense}$ (red), HB$_{\rm
    loose}$ (orange), PD (blue), TH (yellow), B
  (grey).}
\label{fig:ea_20_10}
\end{figure*}

Another interesting feature emerges as we compare the curves of the
order parameters shown in Figure~\ref{fig:ea_2_1} obtained for the
parameters $\left((Q^*)^2,P^*\right) = (2,1)$, with the corresponding
data calculated for the parameter sets $\left((Q^*)^2,P^*\right) =
(0.2,0.1)$ and $\left((Q^*)^2,P^*\right) = (20,10)$, shown in
Figures~\ref{fig:ea_02_01} and~\ref{fig:ea_20_10}, respectively.  The
corresponding curves reveal striking similarities, suggesting that
appropriate scaling relations of the order parameters via the values of the 
quadrupole moment and the pressure hold for the respective ground
states. In contrast (and interestingly), the enthalpy curves obtained
for the different sets of data differ rather substantially in
magnitude. We will discuss a possible background scenario of these
observations in more detail in Appendix~\ref{sec:Reduction of
  parameter space for hard particles}. There we will show that indeed
a scaling relation for the enthalpy of a systems of {\em hard},
quadrupolar particles can be derived (see Appendix~\ref{sec:Reduction
  of parameter space for hard particles: Ground state}). However, it
seems that the softness of our particles -- remember that we consider
in this contribution particles with a {\it soft} (albeit rather steep)
core -- leads to a breakdown of this scaling law. This feature is
presumably due to the fact that a simultaneous scaling of the
quadrupole moment and of the pressure also induces a change of the
density. To take into account this effect properly, we suggest in
Appendix~\ref{sec:Phenomenological enthalpy scaling for non-hard
  particles at ground state} an empirical scaling law for the ground
state enthalpy of a system of soft ellipsoidal particles and provide
numerical evidence for its justification.
\subsection{Results at finite temperatures}
\label{subsec:finite_temperatures}			

\begin{figure*}[htbp]
\begin{center}
\includegraphics[width=13cm]{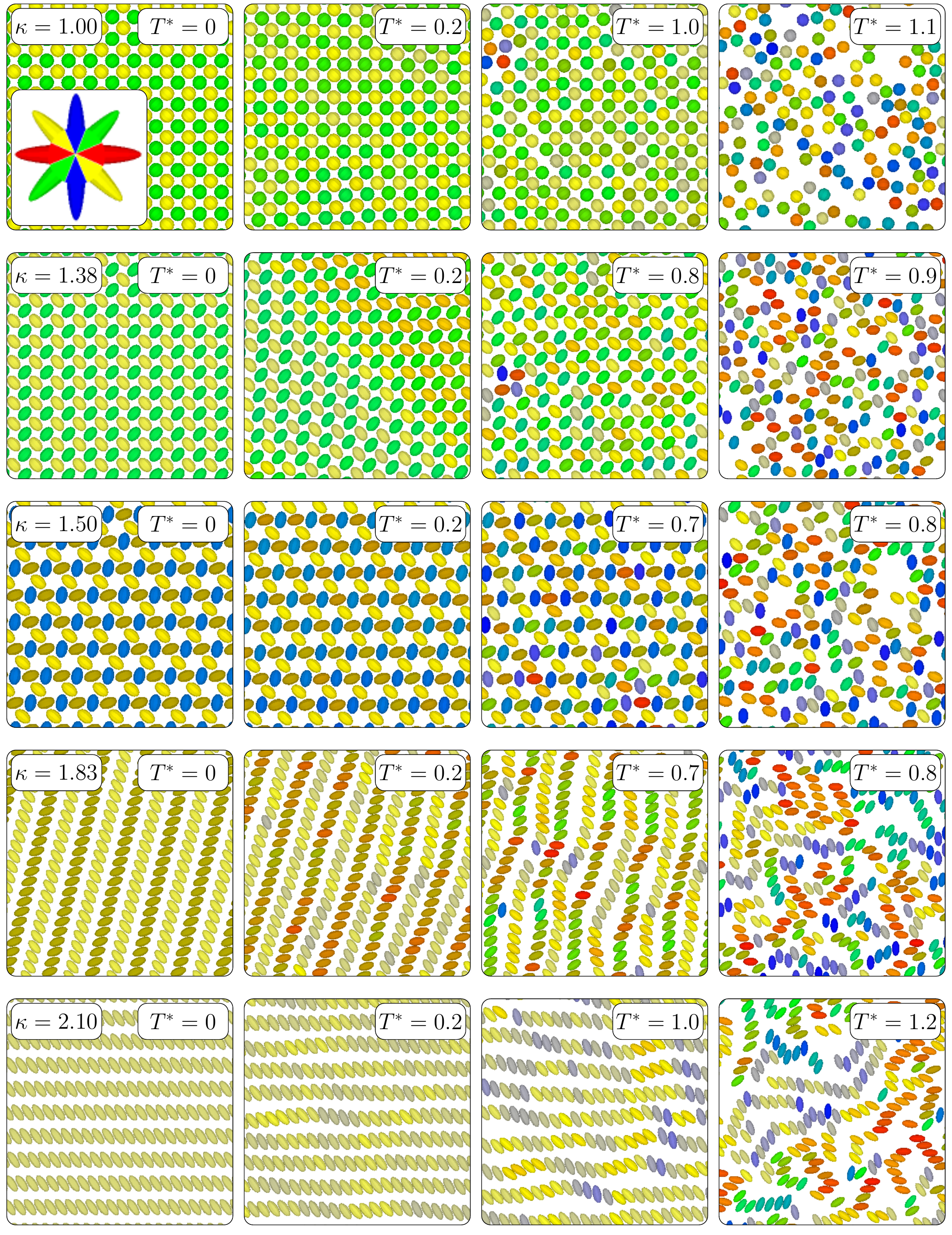}
\caption{Snapshots of equilibrated particle configurations as obtained
  in MD simulations for different values of
  $\kappa=\sigma_{\parallel}/\sigma_{\perp}$ (along rows; as labeled)
  and $T^*=k_{\rm B} T/\epsilon_0$ (along columns; as
  labeled). Different colors indicate different particle orientations
  (see colour scheme in the inset of the top left panel). First row:
  $\kappa=1$ (\textrm{T}$_{\rm sq}$-configuration); second row:
  $\kappa=1.38$ (\textrm{HB}$_{\rm dense}$-configuration); third row:
  $\kappa=1.5$ (\textrm{TH}-configuration); fourth row: $\kappa=1.83$
  (\textrm{HB}$_{\rm loose}$-configuration); fifth row: $\kappa=2.1$
  (\textrm{P}-configuration). Configurations for $T^* = 0$ have been
  obtained via the EA route.}
\label{fig:systems_quadrupolar_mdsnap}
\end{center}
\end{figure*}

In the current subsection, we analyze the MD results that we have
obtained for all considered aspect ratios $\kappa$ and a set of
reduced temperatures $T^*$.  As previously mentioned, the ground state
configurations that we have obtained with the help of the EA
algorithm for a variety of aspect ratios, serve now as initial
configurations for the MD simulations.  Corresponding snapshots for
the most interesting and most representative configurations, obtained
after equilibrating the system, are presented in
Figure~\ref{fig:systems_quadrupolar_mdsnap}. 
    
Before starting the investigations, it is worth to briefly consider typical values of $(Q^*)^2$, $P^*$ in experimental systems.
As an example, we consider the quadrupolar Gay-Berne model of benzene proposed by Golubkov {\it et al.}~\cite{golubkov_2006}.
We use their quadrupole strength, $Q_{\rm benzene}=-30.5812\cdot 10^{-40} {\rm C m}^2$, and their spherical diameter, $\sigma_0\approx 0.307~{\rm nm}$ (the latter value is based on the Gay-Berne contact distance). 
Using the surface tension of water (the 2D equivalent of pressure) defined in Ref.~\onlinecite{vargaftik_1983}, $P_{\rm water}=71.99~10^{-3}\,{\rm N} {\rm m}^{-1}$ at $25^\circ{}\rm{C}$, we arrive at the ratio 
$$(Q_{\rm benzene}^*)^2/P_{\rm water}^*=Q_{\rm benzene}^2/(4\pi \varepsilon_{\rm pm} \sigma_0^4 P_{\rm water})\approx 4.5\text{.}$$

However, since the point quadrupole approximation is known to overestimate the interaction strength for narrow interparticle configurations~\cite{miller_1984,golubkov_2006,heinemann_2015} due to the singularity in $V_{\rm lr}$ [see Eq.~\eqref{eqn:Vlr}], we consider here a reduced value of $(Q^*)^2/P^*=2$ instead of $4.5$.
With this choice we hope to cover not only benzene molecules, but also other quadrupolar molecules.
By identifying  $P^*=1$ with $P_{\rm water}$, we arrive at the following energy scale $\epsilon_0=P_{\rm water} \sigma_0^2/P^*=P_{\rm water} \sigma_0^2=4.09\,{\rm kJ}/{\rm mol}$.
In the following, we present MD simulation results for the parameter pairs $\left((Q^*)^2,P^*\right)=(2,1)$, and $(4,2)$.
We start with a detailed investigation of the structure at $(Q^*)^2=2$, $P^*=1$ and different temperatures.

The MD simulations reveal that the structures predicted for vanishing
temperature remain stable also at low temperature ($T^*=k_{\rm B}
T/\epsilon_0=0.2$, see panels in the second column of Figure
\ref{fig:systems_quadrupolar_mdsnap}). As $T^*$ increases
monotonously, defects start to form (see panels in the third column of
Figure \ref{fig:systems_quadrupolar_mdsnap}): the most common of these
is a slight wave-like modulation of previously straight
lines. Finally, once the temperature has been raised above a certain
threshold value (see panels in the fourth column of Figure
\ref{fig:systems_quadrupolar_mdsnap}), the crystalline order is
rapidly lost. The corresponding transition temperatures depend
strongly on $\kappa$: configurations with $\kappa \approx 1$ (T$_{\rm
  sq}$-configuration) and $\kappa$ far from unity (PD-configuration)
turn out to be the most stable ones: in this situation, particles can
approach very closely and exert thus a strongly attractive
quadrupole-quadrupole interaction; these ordered structures break up
only at temperatures as high as $T^*=1.1$ and $T^*=1.2$,
respectively. In contrast, the rather complicated B-configurations melt
already at temperatures as low as $T^*=0.5$ (with $\kappa=1.66$).

\begin{figure*}[htbp]
\begin{center}
\includegraphics[width=16cm]{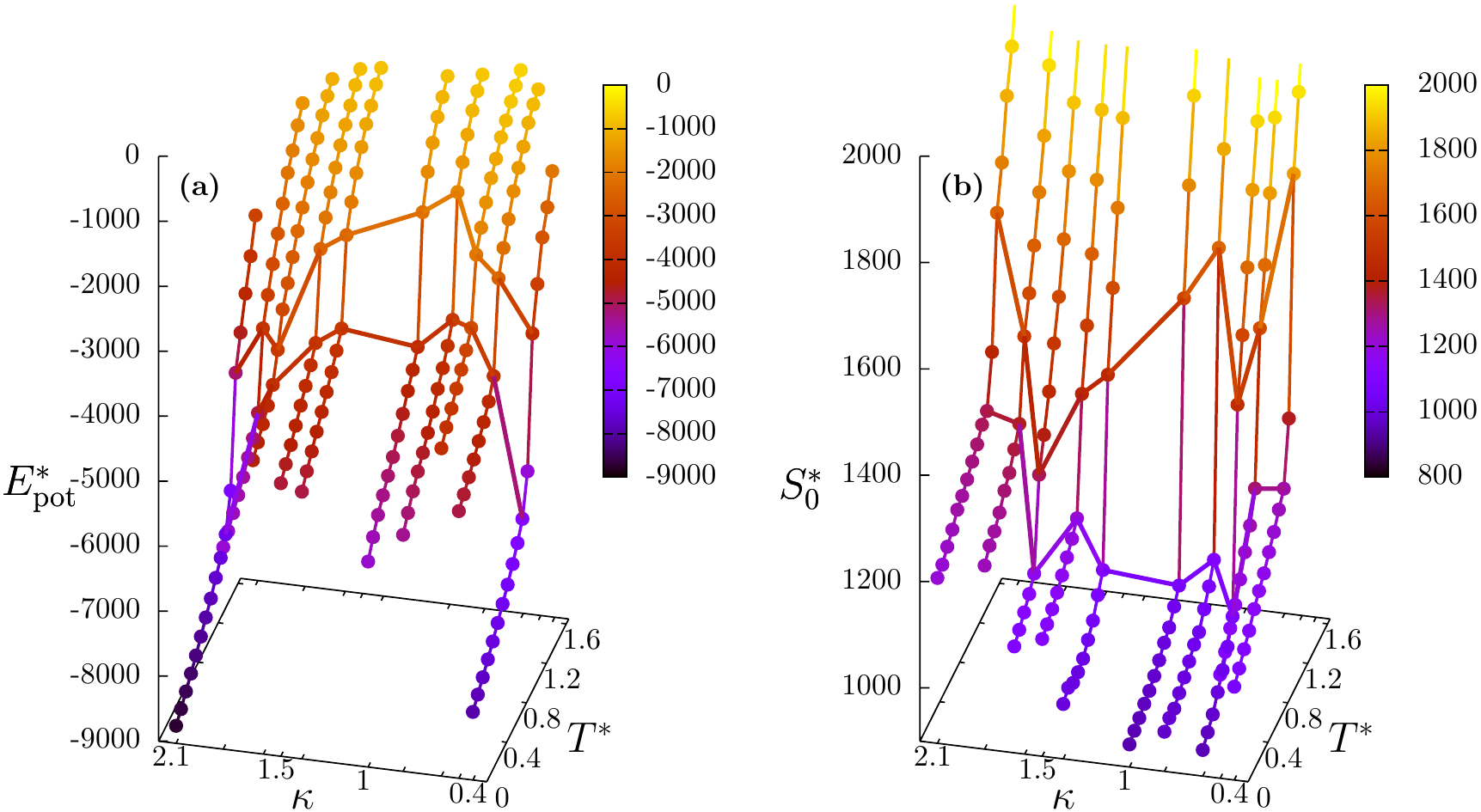}
\end{center}
\caption{Reduced, dimensionless potential energy, $E_{\rm pot}^*$
  (left panel), and dimensionless system area, $S_0^*$ (right panel), as functions
  of the temperature, $T^*$, and of the aspect ratio, $\kappa$.  The
  color code of these quantities is shown on the right hand side of
  each panel. Each symbols represents a state point investigated in an
  MD run; symbols are connected by lines as guides to the eye.  The
  contour lines are explained in the text.}
\label{fig:upotentialvolume}
\end{figure*}

In order to investigate and to locate the transition of the system
from the ordered to the disordered regime, we focus in the following
on the reduced potential energy, $E_{\rm pot}^*$, and the reduced
system area, $S_0^*$, (in units of $\epsilon_0$ and $\sigma_0$,
respectively).  Figure~\ref{fig:upotentialvolume} depicts $E_{\rm
  pot}^*$ and $S_0^*$ as functions of the reduced temperature $T^*$
for all considered values of the aspect ratio $\kappa$. Upon
increasing the temperature, $E_{\rm pot}^*$ progressively decreases in
magnitude and finally approaches zero, reflecting the diminishing role
of particle interactions. At the same time, the area $S_0^*$
increases. 
Interestingly, we observe along this process that both $E_{\rm
  pot}^*(T^{*})$ and $S_0^*(T^{*})$ show discontinuous changes within
very small temperature intervals for all $\kappa$-values investigated;
these ``jumps'' are found for both quantities at approximately the
same temperature.  We attribute these observations to the occurrence
of a first order phase transition. The temperatures that delimit these
intervals are marked in Figure~\ref{fig:upotentialvolume} by contour
lines as functions of $\kappa$.  Only for $\kappa \simeq 1.66$ no such
discontinuity in $E_{\rm pot}^*(T^{*})$ could
be resolved; this fact can be related to the finite size of the
temperature grid.
Intentionally, we have not evaluated the susceptibility, since it is well established
that energy fluctuations are not correctly reproduced within the
Berendsen scheme.~\cite{gromacs-manual,allen_1989}

\begin{figure*}[htbp]
 \centering
 \includegraphics[width=13.5cm]{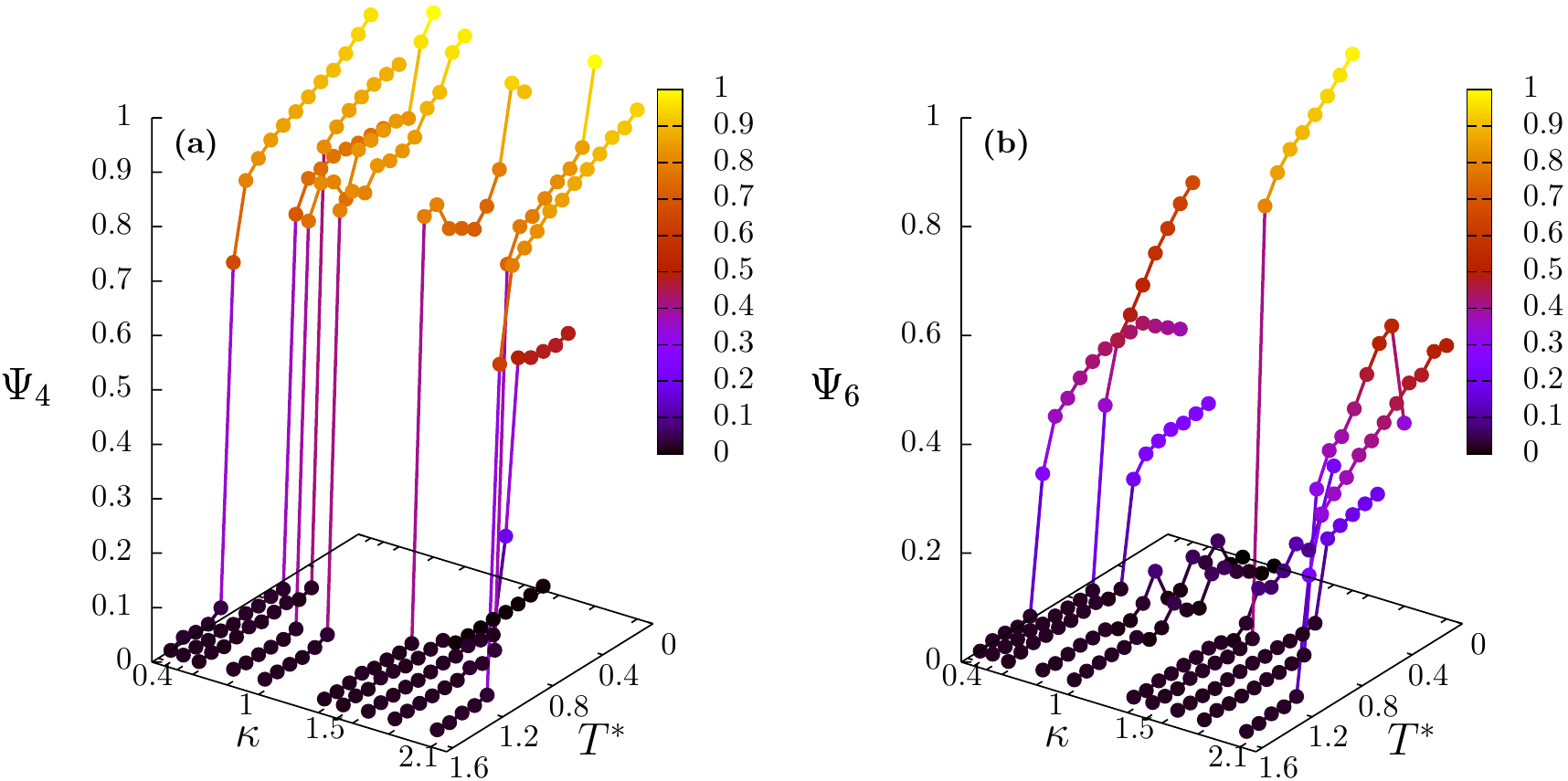}
 \caption{Bond orientational order parameters $\Psi_4$ (panel (a)) and
   $\Psi_6$ (panel (b)) as functions of $\kappa$ for different
   temperatures $T^*$. The color code of these quantities is shown on
   the right hand side of each panel. Each symbols represents a state
   point investigated in an MD run; symbols are connected by lines as
   guides to the eye.}
 \label{fig:psi4psi6}
\end{figure*}

We now arrive at the discussion of the previously defined BOOPs
$\Psi_4$ and $\Psi_6$ [see Eq.~\eqref{eq:Psi}]; they are displayed in
Figure~\ref{fig:psi4psi6}. As already shown in Figure
\ref{fig:ea_2_1}, we find for all our ground state configurations
intervals in $\kappa$ where at least one of the two parameters does
not vanish. Considering now the temperature dependence of these
quantities, we observe for all $\kappa$-values investigated a
discontinuous change of both $\Psi_4$ and $\Psi_6$ (if not zero in the
ground state) at exactly the same temperatures where a discontinuous
change in $S_0^*$ was observed.
To analyze the structure of the system at finite temperatures on an
even more quantitative level, we have calculated in addition various
coefficients of the pair distribution function in an expansion in
terms of rotational invariants. To be more specific, we use a
two-dimensional version of the three-dimensional coefficient functions
of the pair distribution function,\cite{giacometti_2010} i.e.
\begin{widetext}
\begin{align}
g^{l_1 l_2 l}(R)=
\left<\frac{S_0}{N (N-1) \, 2\pi R} 
\sum_{i}\sum _{j\neq i} 
\psi^{l_1 l_2 l}(\mathbf{\hat{r}}_{ij},\mathbf{\hat{u}}_i,\mathbf{\hat{u}}_j) \,  
\delta(\left|\mathbf{r}_{ij}\right|-R) \right>\text{,}
\end{align}
\end{widetext}
where the $\psi^{l_1 l_2 l}(\mathbf{\hat{r}}_{ij}, \mathbf{\hat{u}}_i,
\mathbf{\hat{u}}_j)$ are rotational invariants. Here we focus on the
coefficients $g^{000}(R)$, $g^{220}(R)$, and $g^{202}(R)$ defined via
the following two-dimensional rotational invariants: 
\begin{equation} 
\psi^{000}=1 ~~~~
\psi^{220}=2 (\mathbf{\hat{u}}_i \cdot \mathbf{\hat{u}}_{j})^2 -1 ~~~~
\psi^{202}=2 (\mathbf{\hat{u}}_i \cdot \mathbf{\hat{r}}_{ij})^2 -1.
\end{equation}
The function $g^{000}(R)$ corresponds to the familiar pair correlation function.  The other functions, $g^{202}(R)$ and
$g^{220}(R)$, describe, in addition, a local orientational order of
the particles.  Similar to 3D systems~\cite{kirchhoff_1996} we can
interpret these coefficients as follows: $g^{220}(R)$ provides
information about the conditional probability density of a particle
(relative to the bulk probability density) whose orientation axis is
aligned in parallel (positive value) or orthogonal (negative value) to
the orientation axis of a considered particle; $g^{202}(R)$ describes
the conditional probability density of a particle (again, relative to
the bulk probability density) positioned along or aside the axis of a
considered particle.
		
\begin{figure}[htbp]
\centering
\includegraphics[width=8.5cm]{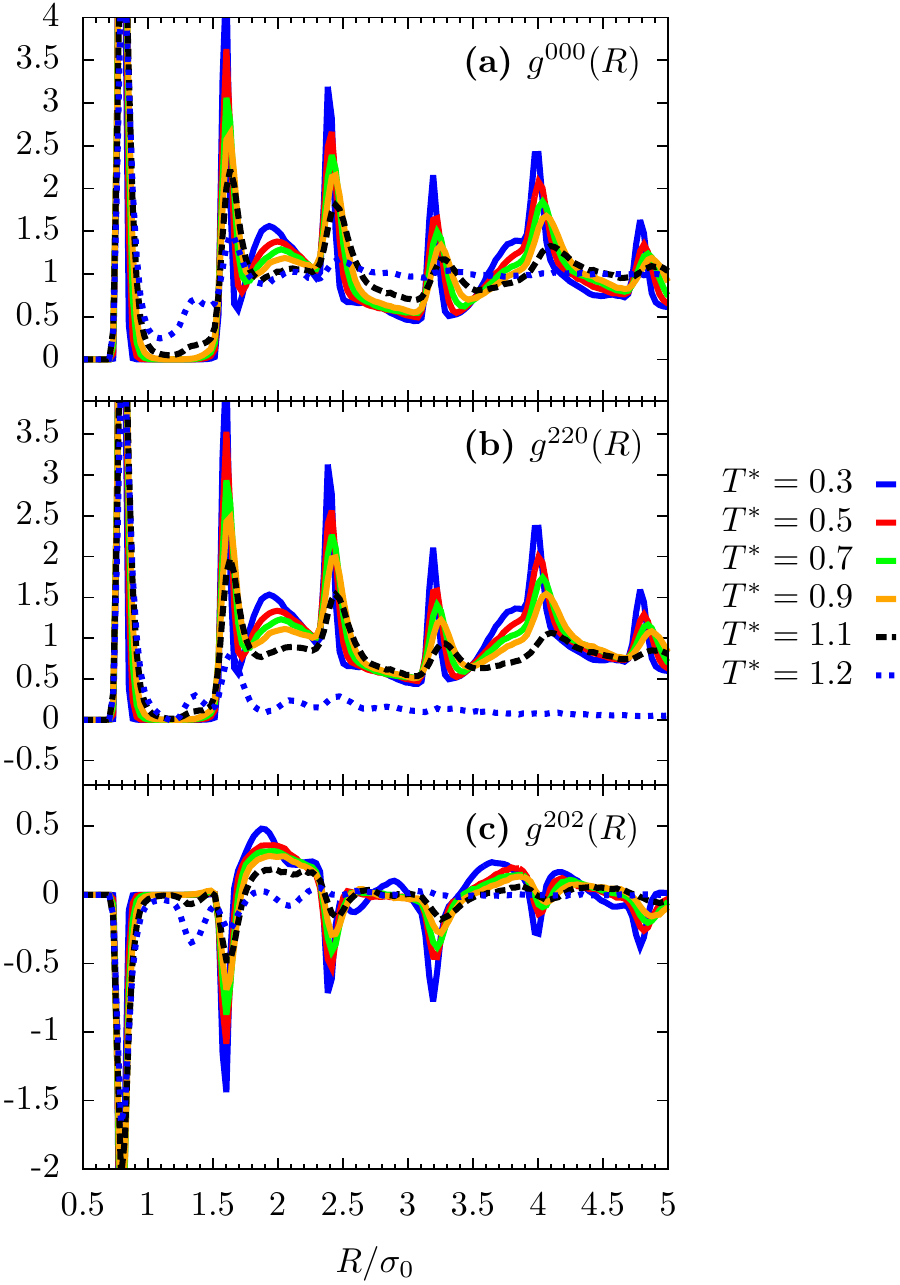}
\caption{Correlation functions $g^{000}(R)$ -- panel (a), $g^{220}(R)$
  -- panel (b), and $g^{202}(R)$ -- panel (c) for $\kappa=2.1$ and
  various temperatures (as labeled).}
\label{fig:gxxxofrk21}
\end{figure}

In the panels of Figure~\ref{fig:gxxxofrk21}, we present numerical
results for these three functions for $\kappa=2.1$ (see also the
corresponding snapshots shown in the panels in the bottom row of
Figure~\ref{fig:systems_quadrupolar_mdsnap}).  For all three
correlation functions the first peak is located at around
$0.8\,\sigma_0$: this position does not exactly mark the face-to-face
alignment of the particles (which occurs at $\sim 0.69\,\sigma_0$) and
thus provides evidence for a slightly shifted parallel configuration,
induced by the quadrupole (see snapshots shown in the panels of the
bottom row in Figure~\ref{fig:systems_quadrupolar_mdsnap}).
For $T^*=1.2$, we observe a rapid decay of $g^{000}(R)$ towards unity
for large particle distances $R$: the crystalline order has completely
vanished, reflected by the missing peaks for larger $R$-values.
Since at low temperatures all particles are oriented in the same
direction, $g^{220}(R)$ is equivalent to $g^{000}(R)$; however, as the
temperature attains $T^*=1.2$, we observe that for large $R$-values the orientational order
is lost and thus $g^{220}(R)$ vanishes already at
$R \simeq 2 \sigma_0$.
Finally, $g^{202}(R)$ oscillates around zero in the ordered phase for
low temperatures ($T^*<1.2$) due to the fact that $g^{202}(R)$ is --
by definition -- not able to contribute to the overall particle
density. For $T^* \gtrsim 1.2$, $g^{202}(R)$ completely vanishes for
large $R$-values ($R \gtrsim 2 \sigma_0$): obviously, a loss of
orientational order occurs for $T^* \gtrsim 1.2$. This temperature
threshold, which can be interpreted as a melting temperature, agrees
fairly well with the temperature where the discontinuous changes in
$E_{\rm pot}^*(T^{*})$ and $S_0^*(T^{*})$ (see
Figure~\ref{fig:upotentialvolume}) and in the BOOPs (see
Figure~\ref{fig:psi4psi6}) were observed. Concluding, we note that
related investigations carried out for other $\kappa$-values led to
analogous conclusions about the corresponding melting temperature.

\begin{figure}[htbp]
\centering
\includegraphics[width=8.5cm]{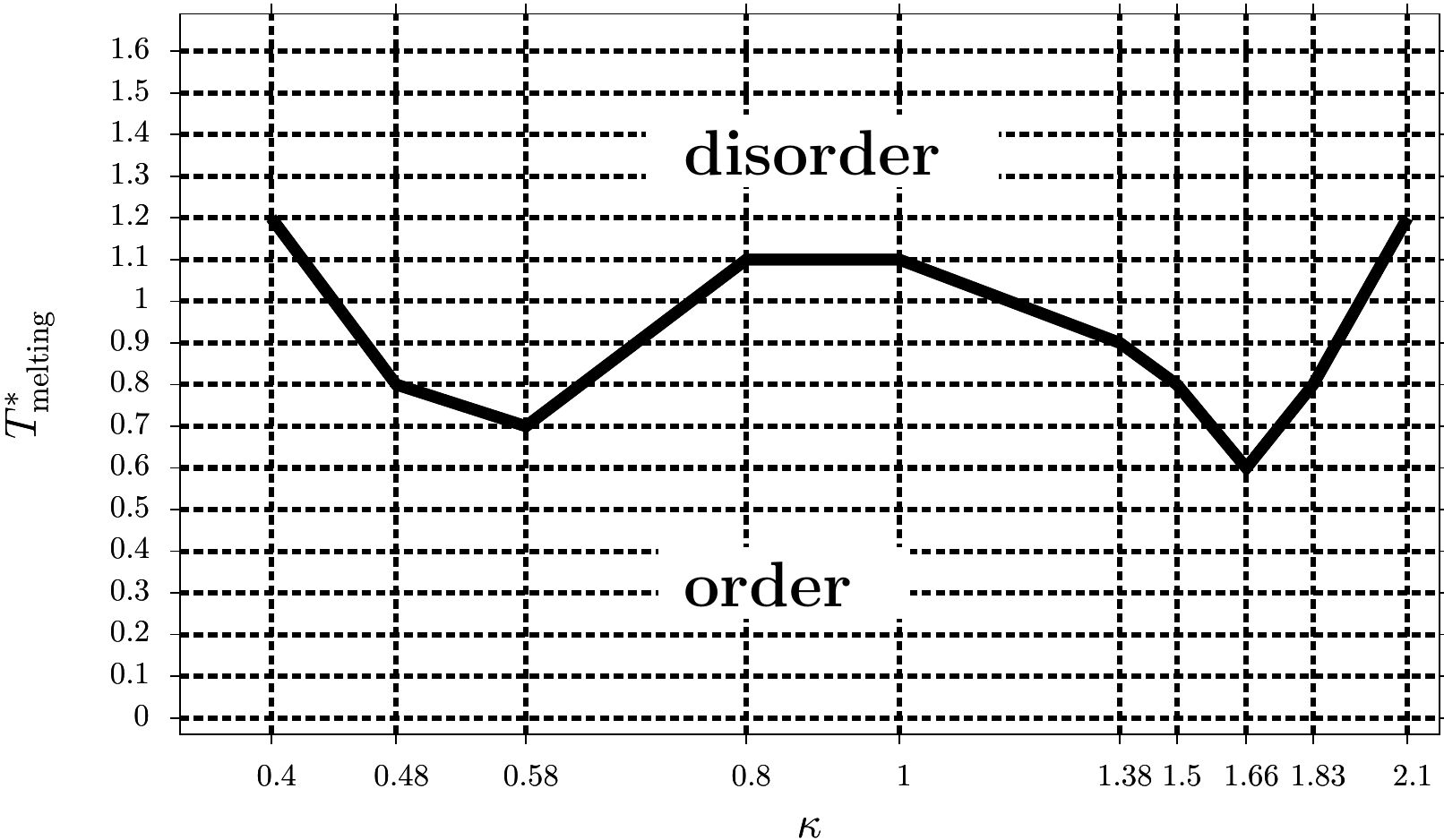}
\caption{Estimate of the melting curve $T^*_{\rm
    melting}((Q^*)^2\!\!=\!\!2,P^*\!\!=\!\!1,\kappa)$ that separates
  the ordered phase (at low temperatures) from the disordered phase
  (at high temperatures); for details cf. text.  Data along on the
  line correspond to results obtained for disordered state
  points.}
 \label{fig:meltingdiagramp1qsqrt2}
\end{figure}

These observations motivate investigations on a {\it melting curve} that separates the
ordered from the disordered phase as we increase the temperature.
This line is displayed in Figure~\ref{fig:meltingdiagramp1qsqrt2} for
all considered aspect ratios $\kappa$.  We emphasize that these data
represent only an estimate for the true, two-phase coexistence lines
characterizing a first-order transition. For the latter, one would
also expect the occurrence of a hysteresis, i.e., the observation of two
different curves, depending on whether the system is heated up or
cooled down from a low or a high-temperature state, respectively. We
briefly come back to this issue below.
Another interesting feature of the melting curve that can be observed
is that it exhibits some degree of symmetry in shape when exchanging
$\kappa$ and $1/\kappa$ (see Figure~\ref{fig:meltingdiagramp1qsqrt2}).
Of course, we would not expect full symmetry since the electrostatic
properties for the cases $\kappa$ and $1/\kappa$ are different (see
bottom panels in Figure~\ref{fig:model}). From the data we can conclude
that for particles with large eccentricities the melting occurs at
higher temperatures than for $\kappa$-values close to unity.

Given the large amount of numerical calculations required to construct
the melting line in Figure~\ref{fig:meltingdiagramp1qsqrt2}, which was
calculated for one particular set of parameters, $(P^{*},(Q^{*})^2)$,
it would be obviously desirable and helpful to have a scaling relation
at hand which allows to easily obtain (or to extrapolate)
corresponding melting lines for other parameter sets. In
Appendix~\ref{sec:Reduction of parameter space in the hard limit at
  finite temperature} we show that such a relation does indeed exist
for {\it hard} particles. This relation states that the probability to
encounter a microscopic configuration of the many-particle system in
phase space is invariant under the simultaneous transformations
$Q^2\rightarrow \mu Q^2$, $P\rightarrow \mu P$, $T\rightarrow \mu T$,
with $\mu$ being a scaling factor. However, in this contribution we
consider {\it soft} particles (even though characterized by a rather
harsh repulsion, see Eq.~\eqref{eqn:Vsr}). Nevertheless, as we discuss
in Appendix~\ref{sec:Scrutinizing the melting curve}, it is also
possible for the system at hand to provide via a suitably adapted
scaling law a rough estimate of the location of the melting line at
scaled parameters. 

Finally and for the sake of completeness, we now discuss our
investigations on the melting transition as obtained in a cooling
process (``simulated annealing'').  The central question that we
address is whether the ground state structures can be reproduced -- at least locally -- with MD simulations starting from a disordered phase
at higher temperatures.
To this end we performed simulations at $(Q^*)^2=2$ and several values of $\kappa$,
initializing the system with random particle positions and
orientations and cooling it down gradually.  The initial pressure and
temperature were set to $P^*=10$ and $T^*=5$, respectively; the
initial box-shape is quadratic with a side length of $50\sigma_0$.
Within the first $200~000$ MD steps we linearly decreased the pressure
and the temperature down to $P^*=1$ and $T^*=0.1$, respectively.  The
simulations extended in total over $410~000$ MD steps. For the other
simulation parameters we refer the reader to Sec.~\ref{subsec:MD}.
\begin{figure}[htbp]
 \centering
 \includegraphics[width=8.5cm]{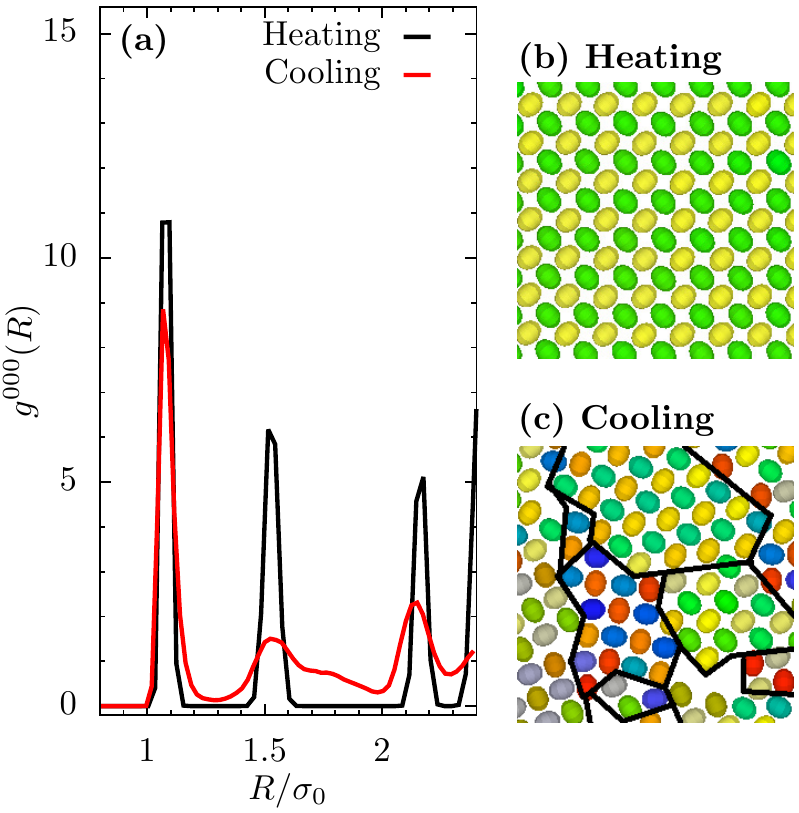}
 \caption{Pair correlation function $g^{000}(R)$ (panel (a)) for a
   system with $\kappa=0.8$ at a temperature $T^*=0.1$, pressure $P^*=1$, and quadrupole strength $(Q^*)^2=2$ after a melting
   (snapshot in panel (b)) and a cooling procedure (snapshot in panel
   (c)). For the color code of the snapshots see inset in
   Figure~\ref{fig:systems_quadrupolar_mdsnap}.}
 \label{fig:k08}
\end{figure}
Our data provide evidence that the predicted ground state could be
obtained via this simulated annealing process only for very few state
points.  In general, the formation of ordered structures via such a
process is hampered and delayed by frustration effects, especially for
$\kappa$-values far from unity, where particles encounter -- due to their
elongated shapes -- difficulties to rotate.  As an example we discuss the
pair correlation function $g^{000}(R)$, obtained for $\kappa=0.8$
(i.e., a value close to unity) and depicted in Figure~\ref{fig:k08}(a).
We observe
  a coincidence in the peak positions of $g^{000}(R)$, but not in their
  heights.  We interpret this deviation as an artefact of the
  crystallite structure appearing after cooling (see
  Figure~\ref{fig:k08}(b) and (c)).  This might be a consequence of
  the lack of long-range order in 2D systems.

\section{Conclusions}
\label{sec:conclusion}

In summary, we have proposed a simple model that mimics the essential
features of elongated, organic molecules without a net charge or
dipole moment: it consists of soft ellipsoids with an embedded linear
quadrupole moment. The preferred orthogonal arrangement of linear
quadrupole moments in close proximity to each other represents an
interesting contrast as compared to dipolar systems, where particles
prefer parallel arrangements, thus often forming chains. The
self-assembly scenarios of our system result from a competition
between the shape anisotropy of the ellipsoids and the quadrupolar
interactions. 

Operating in the NPT ensemble, we have investigated the system for
numerous different sets of system parameters, $(Q^*)^2$ and $P^*$, and
could identify different strategies of the system, depending on the
shape anisotropy $\kappa$. While we always observed configurations of
parallel rows of particles for $\kappa$-values far from unity, the
sequence of structures between these two limiting cases strongly
depends on the competition between the short-ranged and the
long-ranged, electrostatic contributions to the interaction. For
$\kappa \approx 1$ we observe two different configurations with
orthogonal particle arrangement, one of them based on a square
lattice, the other closely related to a hexagonal
lattice. Intermediate ranges of $\kappa$ -- both for $\kappa>1$ as
well as for $\kappa<1$ -- are mostly dominated by two variations of
the herringbone structure. In addition, we observe for selected
$\kappa$-values a non-trivial lattice, closely related to the
trihexagonal tiling. In very small ranges of $\kappa$, more
complicated, branched structures can emerge. However, these turn out
to be in general rather unstable at finite temperatures, as shown in
complementary MD simulations: thus we speculate that they are
metastable at vanishing temperature.
		
For future work, it would be useful to extend our simple model in two
directions: (i) the particles interaction should include a general
quadrupole moment and (ii) the system could be confined to a slab
geometry or be extended to full three dimensions. These extensions
would nicely meet the considerable experimental interest in the
self-assembly of complex organic molecules on surfaces.~\cite{cui_2014,cui_2014_1}
Some of these systems show a remarkable
variety of different structures, which can even be controlled via
external parameters such as magnetic fields~\cite{scheybal_2005} or light with different polarization,~\cite{fang_2010}
opening thereby the route to many interesting technological
applications.~\cite{koch_2008,roales_2012,li_2012,hlawacek_2013,zojer_2000,simbrunner_2013}
The understanding of our simple model considered in the present
contribution could serve as a suitable starting point for related
investigations of the more complex molecules considered in these
studies. Extending the model step-by-step towards more complicated
shapes and to more intricate effective interactions would help to
understand the numerous competing effects.~\cite{heinemann_2014,kleppmann_2015}
		
Finally we note that so-called Inverse Patchy Colloids (IPCs),~\cite{bianchi_2011}
i.e., colloids decorated with charged patches,
represent due to their charge distribution a closely related system,
as they also carry a linear quadrupole moment. IPCs have been observed
to form similar structures as the ones encountered here both in
simulations~\cite{bianchi_2013,bianchi_2014} as well as in
experiments.~\cite{vanoostrum_2015}

\section{Acknowledgements}
This work was supported by the Deutsche Forschungsgemeinschaft (DFG)
within the framework of the CRC 951 (project number A7). 
MA and GK gratefully acknowledge financial support by the Austrian Science
Foundation (FWF) under project numbers P23910-N16 and F41 (SFB ViCoM) and by E-CAM, an
e-infrastructure centre of excellence for software, training, and consultancy in simulation and
modelling funded by the European Union (Proj. No. 676531). Further, financial support from the bilateral projects
PHC-Amadeus-2012 and 2015 (under project numbers 26996UC and 33618YH),
Projekt Amad\'ee (under project numbers FR 10/2012 and FR 04/2015),
the German Academic Exchange Service (DAAD - under project number 10041371), the bilateral projects PHC-Procope-2013 (under project number 28400PD),
and funding by Investissement d'Avenir LabEx PALM (grant
ANR-10-LABX-0039) is acknowledged. 

\appendix

\section{Two particle ground state configurations at particle contact}
\label{sec:Two-particle arrangement with minimal electrostatic energy}

\begin{figure*}
\centering\includegraphics[width=12cm]{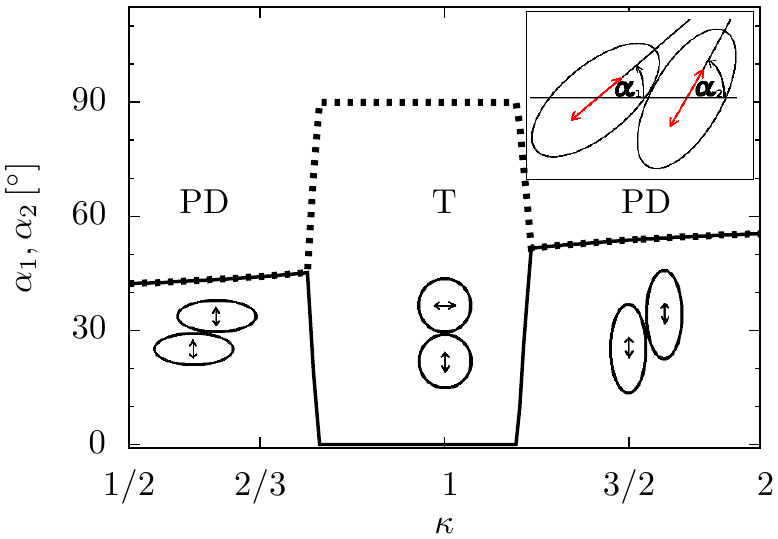}%
\caption{Angles $\alpha_1$ (full line) and $\alpha_2$ (broken line) as
  functions of $\kappa$ that specify the energetically most favorable
  configuration of two quadrupolar particles in direct contact (see
  text and the inset for visualization). Note the logarithmic scale
  along the $\kappa$-axis.}
\label{fig:angles}
\end{figure*}

We consider two quadrupolar particles at contact, i.e.,
$r_{12}=\sigma(\mathbf{\hat{r}}_{12},\mathbf{\hat{u}}_1,\mathbf{\hat{u}}_2)$
[see Eq.~\eqref{eqn:sigma}] and optimize their respective and relative
orientations by minimizing their electrostatic energies. In
Figure~\ref{fig:angles} we display the angles $\alpha_1$ and
$\alpha_2$, enclosed by the vector connecting the centers of the two
particles and the orientations of the respective quadrupolar moments
in the corresponding optimal configurations, as functions of $\kappa$
(for the definitions of the angles see also inset in
Figure~\ref{fig:angles}).  Obviously, the T-configuration is only
stable for values of $\kappa$ close to unity.  For strong anisometry
($\kappa \lesssim 0.75$ or $\kappa \gtrsim 1.2$) the ground state
configuration is parallel displaced (PD).
\section{Reduction of parameter space for hard particles}
\label{sec:Reduction of parameter space for hard particles}

The goal of the present subsection is to derive scaling relations for
the system parameters $Q^2$ 
(quadrupole strength) and $P$ (pressure) under which the thermodynamic
properties of the system (and thus its ground state configurations)
remain unchanged. 
Of course, these relations also hold for our reduced parameters, which we introduced at the end of Sec.~\ref{sec:model}.

We assume in the following a simplified
version of our interaction potential, namely the {\it quadrupolar
  hard-ellipse} (QHE) model, defined by its interaction
\begin{multline}
V_{\rm QHE}(\mathbf{r}_{ij},\mathbf{\hat{u}}_{i},\mathbf{\hat{u}}_{j}; Q^2) 
\\
=
\begin{cases}
V_{\rm lr}(\mathbf{r}_{ij},\mathbf{\hat{u}}_{i},\mathbf{\hat{u}}_{j}; Q^2)&
\mathbf{r}_{ij}>\sigma(\mathbf{r}_{ij},\mathbf{\hat{u}}_i,\mathbf{\hat{u}}_j) \\ 
\infty & \text{else}\end{cases} ;
\label{eqn:quadrupolar hard ellipsoid pot}
\end{multline}
$\sigma(\mathbf{r}_{ij},\mathbf{\hat{u}}_i,\mathbf{\hat{u}}_j)$ has
been defined in Eq. (\ref{eqn:sigma}). Thus, contrary to our original
system (see Sec.~\ref{sec:model}), the QHE model consists of {\it
  impenetrable} ellipsoidal particles, whereas the quadrupolar,
long-range part of the interaction is identical to our original model,
Eq.~\eqref{eqn:Vlr}.

The potential energy, $E_{\rm pot}$, of a specific configuration of
the QHE system is defined via
\begin{align}
E_{\rm pot}(\mathbf{r}^N,\mathbf{\hat{u}}^N; Q^2)& = 
\sum_{i, j; i<j} 
V_{\rm QHE}(\mathbf{r}_{ij},\mathbf{\hat{u}}_{i},\mathbf{\hat{u}}_{j}; Q^2) 
\label{eqn:Epot qhe}
\end{align}
where $\mathbf{r}^N=(\mathbf{r}_1, ... , \mathbf{r}_N)$ and
$\mathbf{\hat{u}}^N=(\mathbf{\hat{u}}_1, ... , \mathbf{\hat{u}}_N)$
specify the microscopic configuration of the particles.

\subsection{Ground state (vanishing temperature)}
\label{sec:Reduction of parameter space for hard particles: Ground state}

At vanishing temperature, the particles are positioned and oriented in
a non-degenerate ground state in an NPT ensemble such that the
enthalpy, $H$, is minimal, that is
\begin{align} \label{eqn:enthalpy GS}
H = \text{min}_{\{\mathbf{r}^N,\mathbf{\hat{u}}^N, S_0 \}} \left[
  E_{\rm pot}(\mathbf{r}^N,\mathbf{\hat{u}}^N; Q^2) + P S_0 \right] .
\end{align}
The scaling properties of the QHE interaction potential (via
Eq.~\eqref{eqn:quadrupolar hard ellipsoid pot}), with the quadrupole
strength $Q^2$ induces a scaling property of the entire potential
energy [cf. Eq.~\eqref{eqn:Epot qhe}], i.e., $E_{\rm pot}( \mu Q^2) =
\mu E_{\rm pot}(Q^2)$; henceforward, $\mu$ is a simple scaling
parameter. Since the enthalpy $H$ is a linear combination of $E_{\rm
  pot}$ and $P$, replacing $Q^2$ by $\mu Q^2$ [cf.
Eq.~\eqref{eqn:enthalpy GS}] leads to the same ground state only if
the pressure also scales with $\mu$. Thus, we can conclude that a
particular ground state configuration of the QHE model (specified by
$\{ \mathbf{r}_{\rm GS}^N,\mathbf{\hat{u}}_{\rm GS}^N, S_0^{\rm GS}
\}$), remains invariant under the transformations $Q^2 \rightarrow \mu
Q^2$ and $P \rightarrow \mu P$.

\subsection{Finite temperatures}
\label{sec:Reduction of parameter space in the hard limit at finite temperature}

We next consider the QHE system at $T>0$. The phase space
probability for the occurrence of a microscopic configuration
(specified by $\{\mathbf{r}^N,\mathbf{u}^N \}$) in a phase space
volume of infinitely small extent in an NPT-ensemble is given by
\begin{widetext}
\begin{align}
\rho(\mathbf{r}^N,\mathbf{u}^N, S_0) 
 \mathrm{d}\mathbf{r}^N \,\mathrm{d}\mathbf{\hat{u}}^N \, \mathrm{d}S_0 & =   
\frac{1}{Z_{\rm conf}} 
\exp \left\{ -(k_{\rm B}T)^{-1} \left[E_{\rm pot}(\mathbf{r}^N,\mathbf{\hat{u}}^N; Q^2) +
P S_0  \right] \right\}
\mathrm{d}\mathbf{r}^N \,\mathrm{d}\mathbf{\hat{u}}^N \, \mathrm{d}S_0 , 
\label{eqn:P_coords}
\end{align}
\end{widetext}
where $Z_{\rm conf}$ is the corresponding partition function of the
ensemble.  It is obvious that Eq.~\eqref{eqn:P_coords} is invariant
under the transformations $E_{\rm pot} \rightarrow \mu E_{\rm pot}$,
$P\rightarrow \mu P$, and $T\rightarrow \mu T$. Since we know from
Eqs.~\eqref{eqn:Vlr} and ~\eqref{eqn:Epot qhe} that the potential
energy of non-overlapping, impenetrable particles is proportional to
$Q^2$, i.e.,
\begin{align}
E_{\rm pot} & =
\sum_{ij; i<j} 
V_{\rm QHE}
(\mathbf{r}_{ij},\mathbf{\hat{u}}_{i},\mathbf{\hat{u}}_{j}; Q^2) 
\propto Q^2 ,
\end{align}
we obtain the simple rule, that $\rho(\mathbf{r}^N,\mathbf{u}^N, S_0)$
is invariant under the scaling law $Q^2\rightarrow \mu Q^2$,
$P\rightarrow \mu P$, and $T\rightarrow \mu T$.

Thus, if we have a reference system (specified by $Q^2$, $P$, and $T$)
at hand, we can simply calculate the properties of another system
(index ``new'') via the following procedure. Given the size of the
corresponding molecule, $\sigma_{\rm new}$, its aspect ratio $\kappa$
and the quadrupole strength, $Q_{\rm new}^2$, it is thus possible to
estimate the order-disorder transition temperature for the new system:
we first calculate the reduced quadrupole strength via $(Q^*_{\rm
  new})^2= Q_{\rm new}^2/(4\pi \varepsilon_{\rm pm} \sigma_{\rm new}^5
\epsilon_0)$ from which we obtain the scaling factor $\mu$ from
$\mu=(Q^*_{\rm new}/Q^*)^2$.
The order-disorder temperature and pressure of the new system then
follow from the corresponding quantities of the reference system,
$T^*_{\rm melting}$ and $P$, via $T^*_{\rm melting, new} = \mu
T^*_{\rm melting}$ and $P^*_{\rm new}=\mu P^*$.

\section{Applicability of parameter scaling for soft particles}

We now explore to which extent we can apply the parameter scaling
relations derived for the QHE system to the actual system of soft
particles investigated in the main part of this paper (see
Sec.~\ref{sec:model}).

\subsection{Ground state (vanishing temperature)}
\label{sec:Phenomenological enthalpy scaling for non-hard particles at ground state}

The total potential energy for a specific configuration of soft
quadrupolar ellipsoids is defined through
\begin{multline}
E_{\rm pot}(\mathbf{r}^N,\mathbf{\hat{u}}^N; Q^2) \\ =
\sum_{i,j;i<j} V_{\rm sr}(\mathbf{r}_{ij},\mathbf{\hat{u}}_{i},\mathbf{\hat{u}}_{j})+
V_{\rm lr}(\mathbf{r}_{ij},\mathbf{\hat{u}}_{i},\mathbf{\hat{u}}_{j}; Q^2) .
\label{eqn:Epot soft rep} 
\end{multline}
As pointed out in the discussion of the ground states in
Sec.~\ref{subsec:ground_state}, the order parameters reveal as
functions of $\kappa$ strong similarities when one compares their
values as obtained for the parameter sets
$\left((Q^*)^2,P^*\right)=(0.2,0.1)$ and
$\left((Q^*)^2,P^*\right)=(20,10)$. This suggests that the
soft particle system fulfills at $T=0$ a similar scaling relation as
the QHE system.  

However, the corresponding enthalpy curves match only in terms of
shape, but not in terms of magnitude.  We thus introduce a {\it
  phenomenological} correction to the simple scaling law of the
enthalpy as developed in Appendix~\ref{sec:Reduction of parameter
  space for hard particles: Ground state}.  Specifically, we assume
that a scaling of $P$ and $Q^2$ with a factor $\mu$ results in a small
rescaling of the ground state coordinates, i.e. $\mathbf{r}_i
\rightarrow \gamma \mathbf{r}_i$, where $\gamma$ attains values close
to unity. We further assume that, when passing from one ground state,
obtained for parameters $(Q^2,P)$ and a filling fraction $\eta$, to a
new ground system (index ``new''), specified by parameters $(\mu Q^2,
\mu P)$ and a filling fraction $\eta_{\rm new}$, the rescaling factor
$\gamma$ is related to the change in density according to
\begin{align}
\gamma=\sqrt{\frac{\eta}{\eta_{\rm new}}} .
\end{align}
As a consequence, the ground state cell volume is scaled with
$\gamma^2$.  

\begin{figure}[htbp]
\begin{center}
\includegraphics[width=8.5cm]{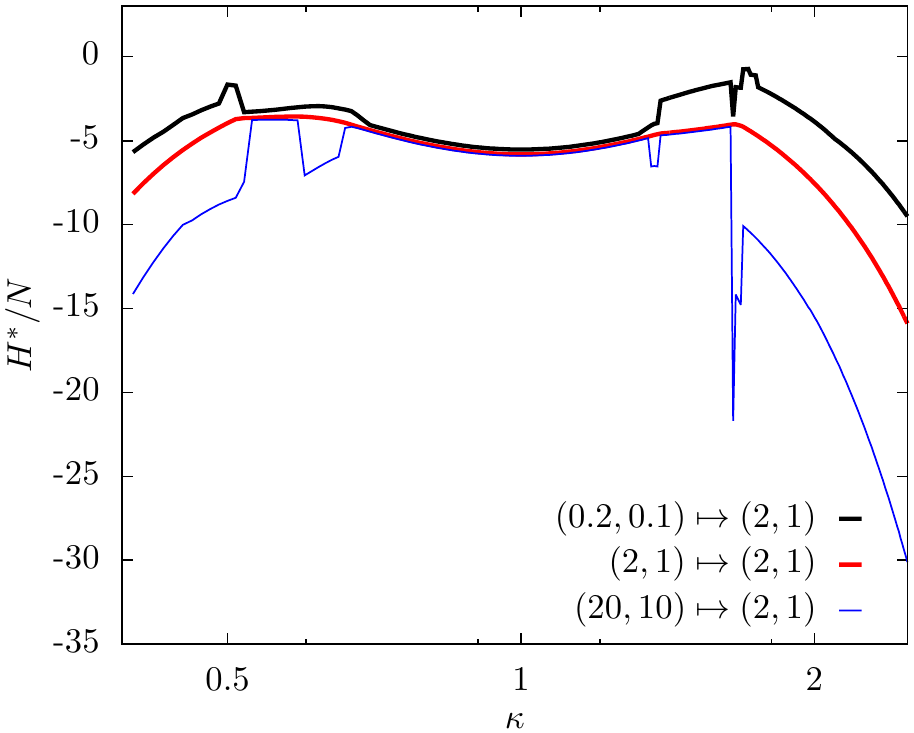}
\end{center}
\caption{Reduced, dimensionless enthalpy curves obtained for the
  parameter sets $\left( (Q^*)^2=0.2, P^*=0.1 \right)$ and $\left(
  (Q^*)^2=20, P^*=10 \right)$, mapped via the scaling laws presented
  in the text (cf. Eq.~\eqref{eqn:law_GS_semi-soft}) onto the enthalpy,
  calculated directly for the set of parameters
  $\left((Q^*)^2,P^*\right)=(2,1)$. }
\label{fig:enthalpy18}
\end{figure}

Regarding the total potential energy, we assume that its
scaling with respect to $Q^2$ and the rescaling of the positions
follows the same law in the QHE system [see Eqs.~\eqref{eqn:Epot qhe},
  ~\eqref{eqn:quadrupolar hard ellipsoid pot} and~\eqref{eqn:Vlr}],
yielding
\begin{align}
E_{\rm pot, new}^{\rm GS}(\gamma \mathbf{r}_{ij},
\mathbf{\hat{u}}_{i},\mathbf{\hat{u}}_{j}; \mu Q^2)
\approx
\frac{\mu}{\gamma^5} 
E_{\rm pot}^{\rm GS}(\mathbf{r}_{ij},\mathbf{\hat{u}}_{i},\mathbf{\hat{u}}_{j}; Q^2)
.
\end{align}
Collecting all contributions, we obtain the following approximate
expression for the enthalpy:
\begin{align}\label{eqn:law_GS_semi-soft}
H_{\rm new}&\approx\frac{\mu}{\gamma^5} E_{\rm pot}^{\rm GS}( Q^2)+
\mu P\,\gamma^2\, S_0^{\rm GS}\nonumber\\
&=
\frac{\mu}{\gamma^5} H+\mu \left[\gamma^2-\frac{1}{\gamma^5}\right]
P\, S_0^{\rm GS} .
\end{align}
We check the applicability of this phenomenological scaling rule by
calculating, as an example, enthalpies of a new system from the values
of the enthalpies obtained at $(Q^*)^2=0.2$, $P^*=0.1$ and
$(Q^*)^2=20$, $P^*=10$. The corresponding values for the enthalpy as
functions of $\kappa$  are presented in Figure~\ref{fig:enthalpy18}.
From these data we can conclude that Eq.~\eqref{eqn:law_GS_semi-soft}
seems applicable for $\kappa$-values close to unity where the
corresponding structures possess a 4-fold symmetry (T-configuration).
For particles with larger aspect ratios, substantial deviations
between the respective enthalpy curves occur.

\subsection{Melting curve}
\label{sec:Scrutinizing the melting curve}

As shown in Appendix~\ref{sec:Reduction of parameter space in the hard
  limit at finite temperature}, the properties of the QHE system
fulfill a parameter scaling relation also at {\it finite} temperature.

\begin{figure}[htbp]
\centering
\includegraphics[width=8.5cm]{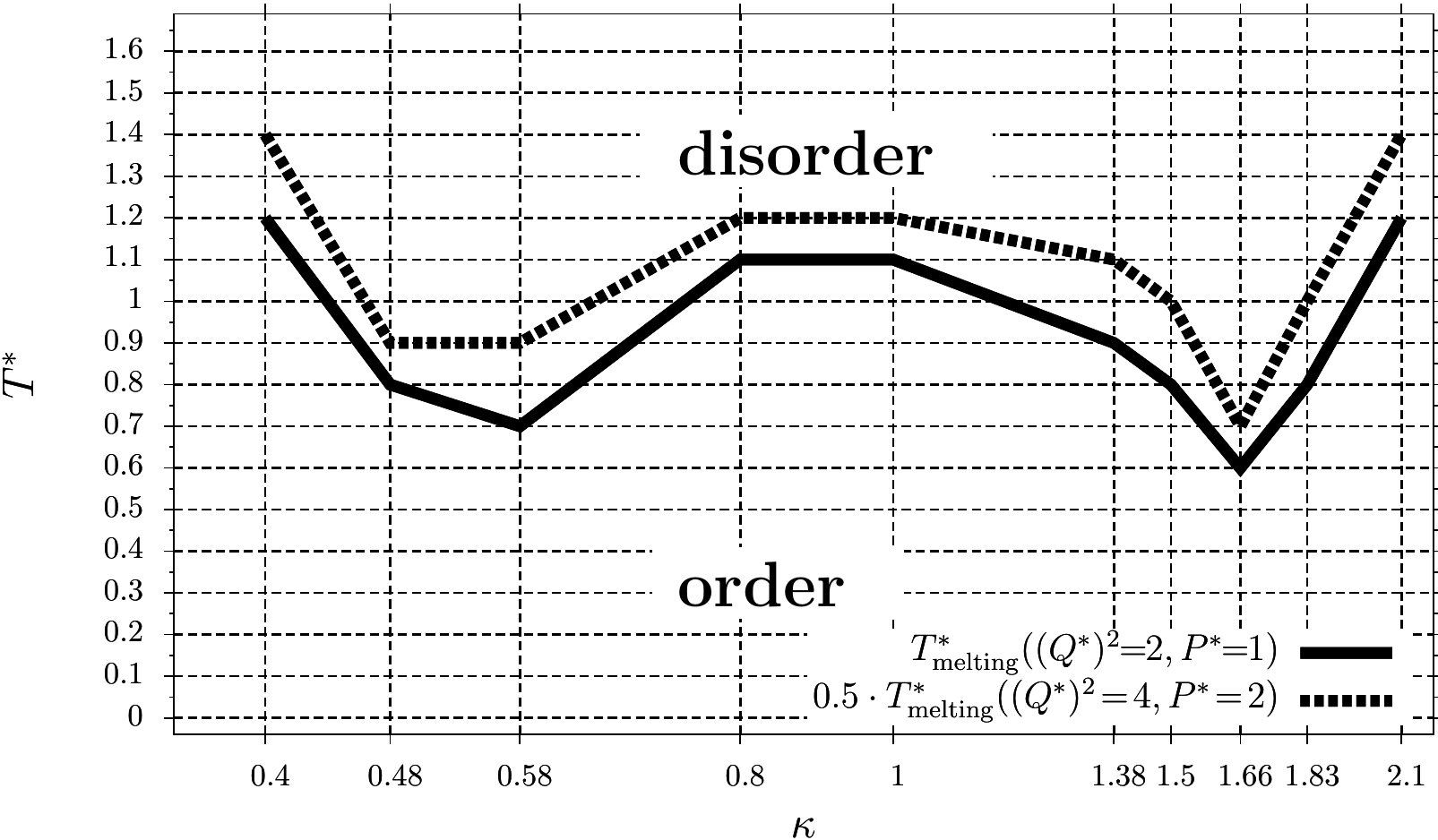}
 \caption{Melting curve $T^*_{\rm melting}((Q^*)^2\!\!=\!\!2,
   P^*\!\!=\!\!1, \kappa)$ that separates the ordered phase (at low
   temperatures) from the disordered phase (at high temperatures). In
   addition, a rescaled melting curve (using $\mu = 2$) for $0.5 \cdot T^*_{\rm melting}((Q^*)^2\!=\!4, P^*\!=\!2, \kappa)$ is shown,
   which has been obtained by applying the scaling laws specified in
   the text.}
\label{fig:meltingdiagram}
\end{figure}

We now consider the question whether this rule can also be extended as
an approximation to a system of soft particles. We are particularly
interested in the melting curve, an example for which is shown in
Figure~\ref{fig:meltingdiagramp1qsqrt2} pertaining to the parameters
$\left((Q^*)^2,P^*\right)=(2,1)$.  To test the scaling rule, we have
carried out the corresponding simulations for the case $\mu=2$,
i.e. $\left((Q^*)^2,P^*\right)=(4,2)$ at temperatures
$T^*=0.2,0.4,\dots,3.2$.  The rescaled melting curve for the new
system, specified by the latter parameter set is displayed in
Figure~\ref{fig:meltingdiagram} in the rescaled form together with the
original melting curve, corresponding to
$\left((Q^*)^2,P^*\right)=(2,1)$.  We observe that the new, rescaled
melting curve is similar in shape to the original one, but the two
sets of data do not coincide: the rescaled curve lies over the entire
$\kappa$-range above the reference curve; we attribute this fact to
the stronger molecular overlaps leading to an increased binding
strength.  Specifically, we observe for all aspect ratios cohesion
energies that are about three times higher for the parameter set
$\left((Q^*)^2,P^*\right)=(4,2)$ at the ground state; the factor three
differs from the expected value of $\mu=2$.  In addition to the
stronger overlaps, also the observed structural archetypes can differ
between the two systems at the ground state. To be more specific we
obtained for $\kappa=1.38$, a TH-configuration for
$\left((Q^*)^2,P^*\right)=(4,2)$, whereas for
$\left((Q^*)^2,P^*\right)=(2,1)$ a T-configuration is observed. We
conclude that the scaling law put forward in Sec.~\ref{sec:Reduction
  of parameter space in the hard limit at finite temperature} only
provides a rough estimate of the results obtained in actual
calculations.



%

\end{document}